\documentclass[english,authoryear]{svjour3}
\usepackage{mathpazo}
\usepackage[T1]{fontenc}
\usepackage{geometry}
\usepackage{babel}
\usepackage{array}
\usepackage{textcomp}
\usepackage{url}
\usepackage{multirow}
\usepackage{amstext}
\usepackage{amssymb}
\usepackage{graphicx}
\usepackage{setspace}
\usepackage[authoryear]{natbib}
\usepackage[unicode=true]
 {hyperref}
\usepackage{breakurl}

\makeatletter

\providecommand{\tabularnewline}{\\}

\makeatother

\begin{document}

\title{Discussion on the spectral coherence between planetary, solar and
climate oscillations: a reply to some critiques}

\author{Nicola Scafetta$^{1,2}$}

\institute{$^{1}$Meteorological Observatory, Department of Earth Sciences,
Environment and Georesources, University of Naples Federico II, Largo
S. Marcellino 10, 80138 Naples, Italy. $^{2}$Active Cavity Radiometer
Irradiance Monitor (ACRIM) Lab, Coronado, CA 92118, USA. $^{3}$Duke
University, Durham, NC 27708, USA. +1(919) 225-7799; nicola.scafetta@unina.it,
nicola.scafetta@gmail.com}
\maketitle
\begin{abstract}
During the last few years a number of works have proposed that planetary
harmonics regulate solar oscillations. Also the Earth's climate seems
to present a signature of multiple astronomical harmonics. Herein
I address some critiques claiming that planetary harmonics would not
appear in the data. I will show that careful and improved analysis
of the available data do support the planetary theory of solar and
climate variation also in the critiqued cases. In particular, I show
that: (1) high-resolution cosmogenic $^{10}Be$  and $^{14}C$  solar
activity proxy records both during the Holocene and during the Marine
Interglacial Stage 9.3 (MIS 9.3), 325\textendash 336 kyear ago, present
four common spectral peaks (confidence level $\gtrapprox95\%$) at
about 103, 115, 130 and 150 years (this is the frequency band that
generates Maunder and Dalton like grand solar minima) that can be
deduced from a simple solar model based on a generic non-linear coupling
between planetary and solar harmonics; (2) time-frequency analysis
and advanced minimum variance distortion-less response (MVDR) magnitude
squared coherence analysis confirm the existence of persistent astronomical
harmonics in the climate records at the decadal and multidecadal scales
when used with an appropriate window length ($L\approx110$ years)
to guarantee a sufficient spectral resolution to solve at least the
major astronomical harmonics. The optimum theoretical window length
deducible from astronomical considerations alone is, however, $L\gtrapprox178.4$
years because the planetary frequencies are harmonics of such a period.
However, this length is larger than the available 164-year temperature
signal. Thus, the best coherence test can be currently made only using
a single window as long as the temperature instrumental record and
comparing directly the temperature and astronomical spectra as done
in \citet{Scafetta2010} and reconfirmed here. The existence of a
spectral coherence between planetary, solar and climatic oscillations
is confirmed at the following periods: 5.2 year, 5.93 year, 6.62 year,
7.42 year, 9.1 year (main lunar tidal cycle), 10.4 year (related to
the 9.93-10.87-11.86 year solar cycle harmonics), 13.8-15.0 year,
$\sim$20 year, $\sim$30 year and $\sim$61 year, 103 year, 115 year,
130 year, 150 year and about 1000 year. This work responds to the
critiques of \citet{Cauquoin}, who ignored alternative planetary
theories of solar variations, and of \citet{Holm}, who used inadequate
physical and time frequency analysis of the data. 

\textbf{Cite as:} Scafetta, N., 2014. Discussion on the spectral coherence
between planetary, solar and climate oscillations: a reply to some
critiques. Astrophys Space Sci 354, 275-299, 2014. DOI: \href{http://link.springer.com/article/10.1007\%2Fs10509-014-2111-8}{10.1007/s10509-014-2111-8}
\end{abstract}

\maketitle

\section{Introduction}

\citet{Wolf} proposed that solar variability could be modulated by
the combined effect of the planets, in particular by Venus, Earth,
Jupiter and Saturn. However, solar scientists have been skeptical
about the possibility of a planetary theory of solar variation because
according Newtonian classical gravitational physics the planets appear
to be too far from the Sun to effectively influence its activity.
For example, Newtonian tidal accelerations induced by the planets
on the Sun's tachocline appear to be too small \citep[e.g.: ][]{Callebaut,Scafetta2012d}.
However, Newtonian gravitational physics alone does not explain how
the Sun and the heliosphere work because, as well known, electromagnetism,
quantum-mechanics and nuclear fusion physics are required as well
\citep{Bennett}. 

Strong internal nuclear fusion amplification tidal mechanisms \citep{Scafetta2012d,Wolff},
torques acting on a non-spherical tachocline \citep{Abreu} and electromagnetic
coupling throughout the heliosphere \citep[e.g.:][]{ScafettaW2013}
have been proposed as possible mechanisms. 

Evidences for a planet-induced stellar activity have been noted in
numerous stars, in particular when the phenomenon becomes macroscopic
due to the presence of a hot Jupiter \citep{Scharf,Shkolnik2003,Shkolnik2005}.
\citet{Wright} detected a quasi 9-year activity cycle in a star with
a Jupiter-like planet on a 9.2-year circular orbit with radius 4.2
AU.

In the case of our Sun, if its activity is regulated by planetary
harmonics, the problem is expected to be quite subtle. Multiple planetary
harmonics would produce complex beats with the solar internal cycles,
and solar activity would not mirror just a single and easily recognizable
planetary harmonic (e.g. the 11.86-year Jupiter orbital cycle) but
would non-linearly respond to a complex synchronization pattern emerging
from the harmonics of the entire solar system \citep[e.g.:][]{Jakubcova,Scafetta2014}. 

Although the physical problem is evidently still open, numerous recent
works have promoted the theory by claiming that solar activity presents
empirical signatures of planetary harmonics from the monthly to the
millennial time scales \citep[e.g.: ][]{Abreu,Charvatova,Fairbridge1987,Jose,McCracken2013,McCracken2014,Scafetta2012c,Scafetta2012d,ScafettaW2013,Sharp,Shirley,Tan}.
Because the Earth's climate appears closely linked to solar variations
\citep[e.g.:][and many others]{Hoyt1997,Steinhilber}, similar harmonics
have also been found in the climate system at multiple decadal, secular
and millennial time scales \citep[e.g.: ][]{Abreu,Scafetta2010,Scafetta2013b}.
Aurora, sunspot, and meteorite fall records too present planetary
harmonics and suggest gravitational, electromagnetic and luminosity
links between astronomical and climatic records \citep[e.g.: ][]{Scafetta2012a,Scafettaw2013a}. 

Spectral coherence between astronomical and solar records was found
from the monthly to the millennial time scales \citep[e.g.: ][and others]{Scafetta2010,Scafetta2012c,Scafetta2013b}
and, as \citet{Wolf} conjectured, also the very 11-year Schwabe solar
cycle appears mostly synchronized to the harmonics generated by the
orbital revolution of Venus, Earth, Jupiter and Saturn \citep[e.g.: ][]{Hung,Salvador,Scafetta2012c,Scafetta2012d,Wilson}.
A special issue collecting works on ``pattern in solar variability,
their planetary origin and terrestrial impacts'' has been recently
published \citep[and other authors]{Morner,Scafetta2014}. 

However, critiques questioning some of the above claims have also
appeared. Herein I will focus on the critiques by \citet{Cauquoin}
and \citet{Holm}.

\citet{Cauquoin} critiqued \citet{Abreu} arguing that if the solar-planetary
harmonic coherence highlighted by \citet{Abreu} in several $^{14}C$
and $^{10}Be$  series throughout the Holocene period \citep{Steinhilber}
were real, similar harmonics had to be found also in \textit{``the
record of $^{10}Be$  in the EPICA (European Project for Ice Coring
in Antarctica) Dome C (EDC) ice core from Antarctica during the Marine
Interglacial Stage 9.3 (MIS 9.3), 325\textendash 336 kyear ago''}.
About the secular scales \citet{Cauquoin} claimed to have found only
a very \textit{``limited similarity with the periodicities ... predicted
by the planetary tidal model of \citet{Abreu}.''}

\citet{Holm} critiqued \citet{Scafetta2010} by arguing that on the
decadal and multidecadal scales time-frequency analysis based on $L=60$
year windows and magnitude squared coherence analysis based on $L=20$
and 30 year windows would show time-varying spectral lines that look
\textit{``very different from the nearly constant lines in the time-frequency
plot for the speed of the center of mass of the solar system (SCMSS).''}
Both critical studies were interpreted as questioning the planetary
theory of solar and Earth's climate variability. 

However, in addition to Scafetta's findings several solar and climatic
oscillations have been identified throughout the Holocene. These oscillations
can be easily associated with many well-known astronomical harmonics.
\citet{Pustilnik-1} found an influence of the 11-year solar activity
cycle on the state of the wheat market since medieval England. \citet{Chylek2011}
found evidences for a prominent near 20-year oscillation in multisecular
ice-core records. 50 to 70-year climatic oscillations have been discovered
in numerous climatic instrumental and proxy records \citep[e.g.:][and many others]{Agnihotri,Davis,Jevrejeva,Klyashtorin,Knudsen,Loehle,Mazzarella,Qian,Scafetta2010,Scafettahum}.
\citet{Qian,Puetz,Scafetta2012c} found evidences for a quasi 115-year
climatic oscillation in millennial multiproxy climatic records. A
quasi 900-1000 year oscillation has also been observed in numerous
climatic records throughout the Holocene \citep{Bond,Christiansen,Kerr,Scafetta2012a}.
Similar oscillations are found among proxy records of solar activity,
in the aurora records and in the oscillations of the heliosphere \citep[e.g.:][]{Ogurtsov,Jakubcova,Scafetta2012a,ScafettaW2013}.
Another set of climatic oscillations appear to be lunar related such
as the 8.85-9.3 year oscillation found in the Atlantic Multidecadal
Oscillation and in the Pacific Decadal Oscillation indexes \citep[c.f.:][]{Scafetta2012b,Manzi}.
Additional Soli-lunar harmonics are present too and interfere with
each other generating the El Niño\textendash Southern Oscillation\textendash like
inter-annual variation \citep[cf. ][]{Wang2012}. 

In this paper I comment and refute \citet{Cauquoin} and \citet{Holm}'s
critiques, and I upgrade my previous analysis \citep{Scafetta2010}
with more advanced methodologies and arguments. The main astronomical
periods of interest in this work are at about 5.2 years, 5.93 years,
6.62 years, 7.42 years, 9.1 years (main average solar-lunar tide cycle),
10.4 years (related to the 9.93-10.87-11.86 year solar cycle harmonics),
13.8 years, $\sim$20 years, $\sim$30 years and $\sim$61 years,
103 years, 115 years, 130 years, 150 years and quasi 1000 years \citep[cf.: ][]{Scafetta2010,Scafetta2012a,Scafetta2012c,Scafetta2012d,Scafetta2014}.
Other planetary astronomical periods in solar records have been found,
e.g. numerous sub annual oscillations \citep[e.g.: ][]{Shirley,ScafettaW2013,Scafettaw2013a,Tan},
the $\sim$87 and $\sim$207 year oscillations \citep[e.g.: ][]{Abreu,Bond,Scafetta2012c,Scafettaw2013a}
and others, but these oscillations are not discussed in detail here
because they were not analyzed in the critiqued works.

\section{Common evidences for a planetary influence on solar activity 330,000
years ago and during the Holocene}

\citet{Cauquoin} analyzed a record of $^{10}Be$  in the EPICA (European
Project for Ice Coring in Antarctica) Dome C (EDC) ice core from Antarctica
during the Marine Interglacial Stage 9.3 (MIS 9.3), referring to 325\textendash 336
kyear ago. Their $^{10}Be$  records are shown in Figure 1A and their
Fourier spectra are shown in Figure 1B-C. According \citet{Cauquoin}
own figures, their $^{10}Be$  concentration and flux records are
quite noisy and likely effected by a low frequency component that
they detrend with a filtering: see Figure 1A. Only four secular frequencies
were found at about 103 year (99\% confidence), 115 year ($\approx$95\%
confidence), 130 year (99\% confidence) and 150 year ($\approx$95\%
confidence). Note that \citet{Cauquoin} highlighted with dash black
lines only the spectral peaks at the 99\% confidence level ignoring
the other two spectral peaks that can be discerned about at the 95\%
confidence level curve. In Figure 1 I highlighted with red lines the
95\% confidence spectral peaks. 

By taking into account these four spectral peaks the spectral analyses
among the three studied records are consistent to each other: compare
Figure 1B, 1C and 1D. The 95\% confidence level is typically used
in the scientific literature when Fourier or wavelet spectral analysis
are adopted \citep[e.g.: ][http://web.atmos.ucla.edu/tcd//ssa/ ; http://duducosmos.github.io/PIWavelet/]{Ghil2002,Knudsen}.

\begin{figure}[!t]
\begin{centering}
\includegraphics[width=0.8\textwidth]{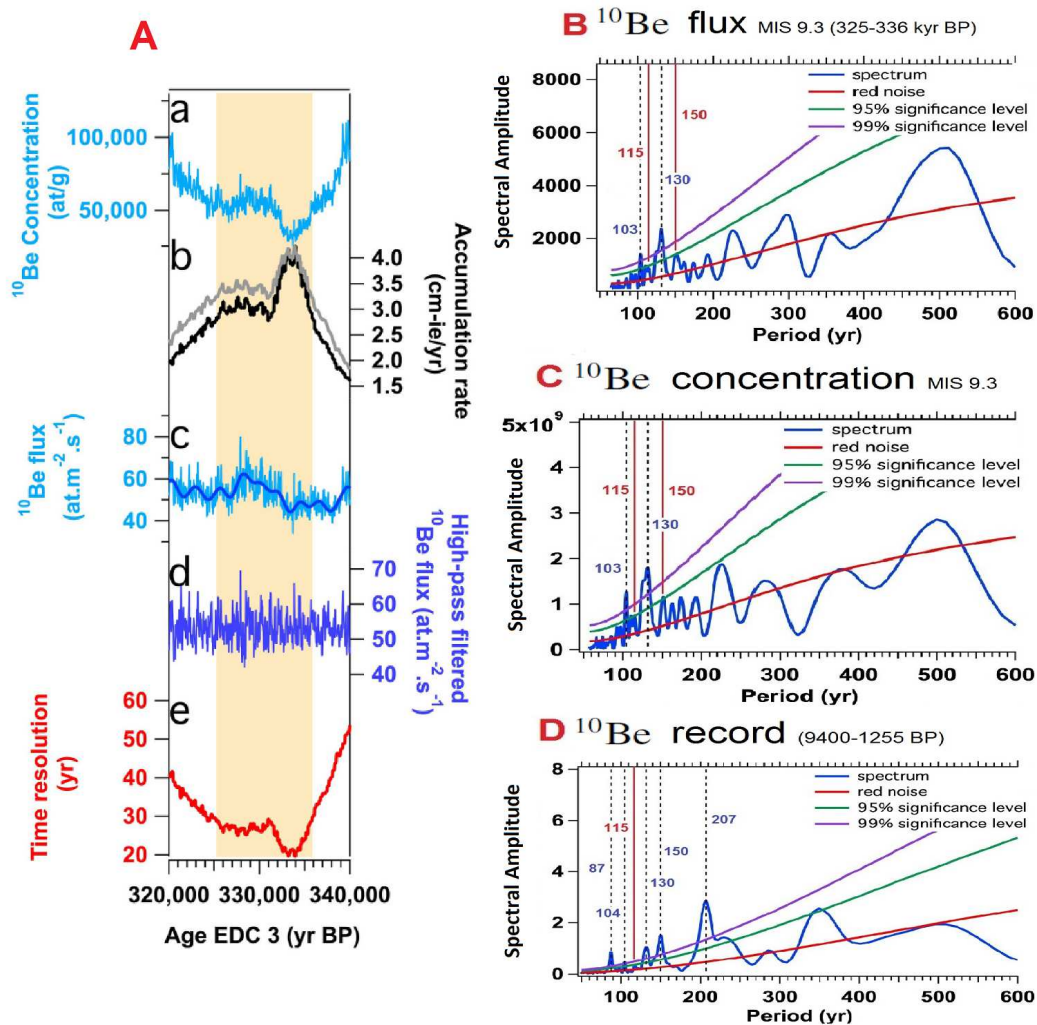}\protect\caption{{[}A{]} $^{10}Be$ concentration (top); $\delta D$ and ice accumulation
rate; $^{10}Be$  flux; 2000-year high-band passed $^{10}Be$  flux;
and $^{10}Be$  sample resolution in EDC ice core (bottom). {[}B,
C, D{]} Fourier spectrum analysis of the $^{10}Be$  flux record at
EDC during the interglacial period MIS 9.3 (325\textendash 336 kyear),
of the $^{10}Be$  concentration profile during MIS 9.3 and of the
composite from \citet{Steinhilber} during the Holocene. The figures
are adapted from \citet{Cauquoin} and highlight 4 main common frequencies
at about 103, 115, 130 and 150 years with a statistical confidence
of $99\%$ (blue) and $\sim95\%$ (red). The Fourier analyses and
their confidence error estimates are based on \citet{Schulz}.}

\par\end{centering}

\end{figure}

\citet{Cauquoin} questioned the planetary theory of solar variation
because they claimed a \textit{``very limited similarity between
the periodicities in this record compared to those found in a proxy
record of solar variability during the Holocene \citep{Steinhilber},
or those predicted by the planetary tidal model of \citet{Abreu}.''}
However, it is evident that \citet{Cauquoin} could not make any definitive
conclusion about periodicities shorted than 100 years and larger than
150 years because according their own confidence level estimates their
records do not have sufficient statistical power in those time scales,
as evident in their own figures reproduced herein in Figures 1B and
1C. 

\citet{Cauquoin} also showed wavelet analysis diagrams using the
methodology proposed by \citet{Grinsted}. These diagrams show an
intermittent patterns. However, wavelet analysis does not substitute
Fourier analysis. Wavelets use small sub-windows whose length is a
certain multiple of the analyzed spectral period, known also as scale.
Because the left and right border confidence lines used in the wavelet
diagrams of \citet{Cauquoin} at period of 1000 years are about 1500-year
wide, the effective wavelet length was about 3 times the spectral
period. As explained in Section 3.4, to separate close harmonics with
spectral analysis methodologies, which include also wavelet based
methodologies, there is a need of using segments longer than the beat
period among contiguous harmonics. The contiguous beat periods among
harmonics with periods of 103, 115, 130 and 150 years are 987, 996
and 975 years respectively (Eq. \ref{eq:beat}). Thus, even if the
records were error-free, which is unrealistic, \citet{Cauquoin} wavelet
analysis can not separate the 4 secular harmonics because the minimum
required window length is 10 times the spectral period while the length
of typical wavelets such as the Morlet or Mexican Hat and others is
up to about 3 times the analyzed scale, as implicit in \citet{Cauquoin}
figures. Thus, \citet{Cauquoin} wavelet analysis can only highlight
the mutual beats among the four frequencies of interest and show intermittent
patterns, as their figures depict.

Figure 1D shows that the Holocene solar proxy model by \citet{Steinhilber}
also presents spectral peaks at about 104 year (99\% confidence),
115 year ($\sim$95\% confidence), 130 year (99\% confidence) and
150 year (99\% confidence) \citep{Abreu,Cauquoin,Scafettaw2013a}.
The model of \citet{Abreu}, which was based on the planetary torque
on the Sun's tachocline, was able only to highlight peaks at about
104 and 150 years. \citet{Poluianov} showed that these peaks could
be aliasing artifacts caused by the calculation algorithm that was
based on an annual average while three planets (Mercury, Venus and
Earth) have orbital periods $\leq1$ year. In any case, \citet{Abreu}
planetary model could explain at most two spectral peaks (at about
103-104 and 150 year periods), but could not explain the other two
observed spectral peaks (at about 115 and 130-year periods). 

However, \citet{Cauquoin} did not realize that the alternative planetary
model proposed by \citet{Scafetta2012c}, which was based on the combination
of the tidal harmonics of Jupiter and Saturn modulating the 11-year
solar cycle plus a non-linear solar and geophysical response producing
the final analyzed nucleotide output records. As shown below, Scafetta's
model predicts major solar harmonics at about 115 and 130 years and
the other harmonics of interest.

As briefly summarized in the Appendix, \citet{Scafetta2012c,Scafetta2014}
noted that spectral analysis of the sunspot records since 1700 reveals
that the Schwabe solar cycle is characterized by three main spectral
peaks at about 9.93, 10.87 and 11.86 years: see Figure 2A. The spectral
peak at 9.93 year is coherent to the spring tidal harmonic generated
by Jupiter and Saturn ($P_{JS}=0.5*P_{J}P_{S}/(P_{S}-P_{J})=9.93$
year, where $P_{J}=11.86$ years and $P_{S}=29.45$ years are the
orbital periods of Jupiter and Saturn, respectively), while the 11.86-year
spectral peak is coherent with the tidal harmonic generated by the
orbital period of Jupiter. The central 10.87 year harmonic is the
central Schwabe solar cycle period \citep[cf. ][]{Pustilnik-1}. The
central $\sim$10.87 harmonic may be generated by the solar dynamo
responding non-linearly to planetary harmonics \citep[e.g.: ][]{Hung,Scafetta2012c,Wilson}.

This result suggests that the sunspot cycle is modulated by the two
Jupiter-Saturn planetary tidal forces. \citet{Scafetta2012d} proposed
a physical model to explain how these tides could effect solar luminosity,
but alternative electromagnetic mechanisms may be present too and
should be characterized by similar and/or related harmonics. 

\begin{figure}[!t]
\centering{}\includegraphics[width=0.8\textwidth]{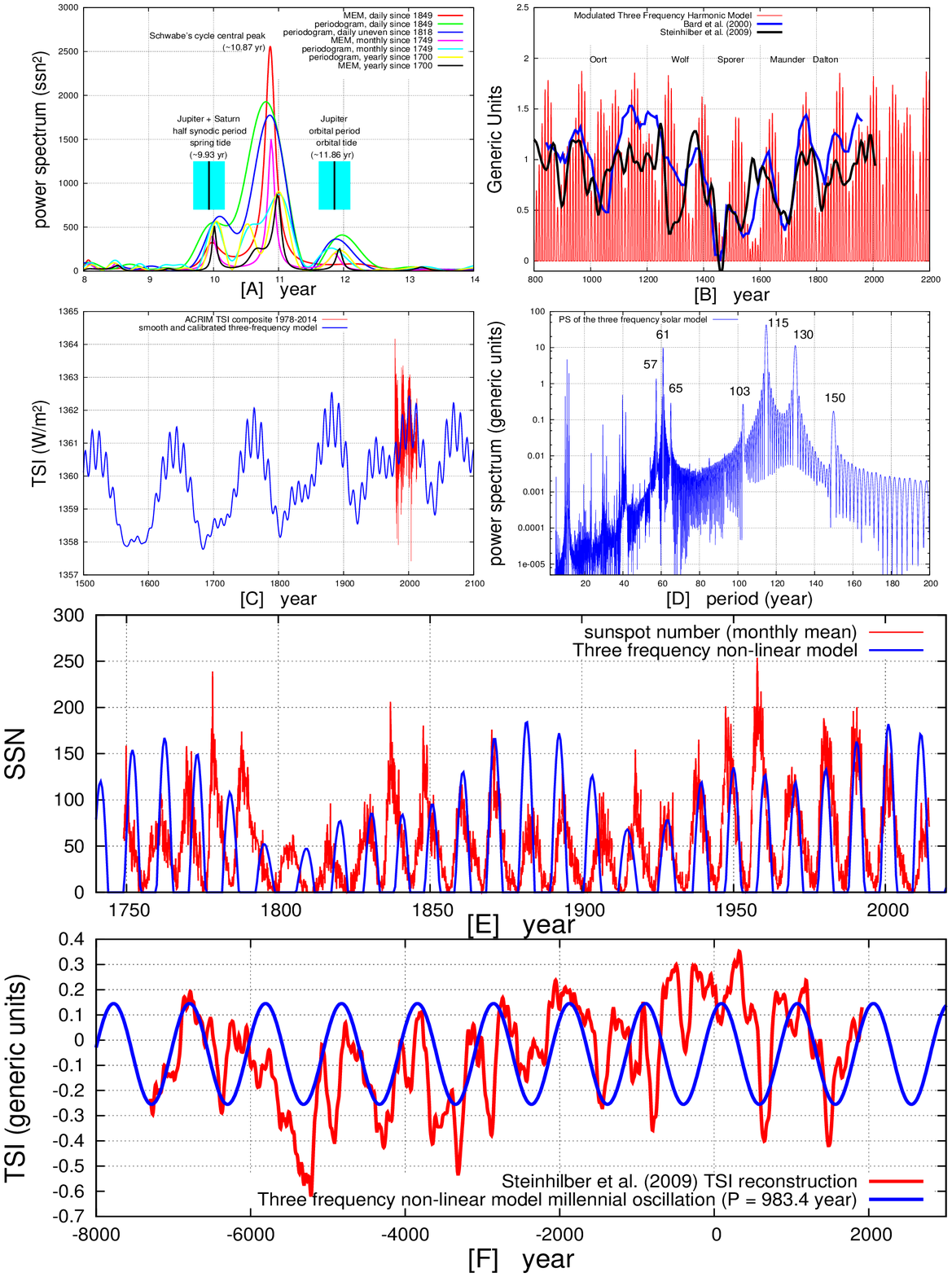}\protect\caption{{[}A{]} Power spectra of sunspot number records and the three-frequency
Schwabe solar cycle \citep[from: ][]{Scafetta2012c,Scafetta2014}.
When the sunspot record since 1700 is used the statistical uncertenty
of the spectral peaks is $\nabla P\approx\pm0.2$ year using Eq. \ref{eq:error}.
{[}B{]} Comparison of the three-frequency non-linear solar model \citep{Scafetta2012c}
and nucleotide solar proxy models \citep{Bard,Steinhilber}. {[}C{]}
An attempted smooth of the three-frequency solar model calibrated
on the ACRIM total solar irradiance satellite record \citep{ScafettaW2014}.
{[}D{]} Periodogram of the three-frequency solar model and main secular
frequencies. Note the four spectral peaks at 103, 115, 130 and 150
years. {[}E{]} Comparison of the three-frequency non-linear solar
model (blue) and the sunspot number record since 1750 (red). {[}F{]}
Comparison of the millennial oscillation predicted by the three-frequency
non-linear solar model (blue) versus the TSI proxy model by \citet{Steinhilber}
(red). }
\end{figure}

\citet{Scafetta2012c} constructed a simple three-frequency solar
model that uses as input a simple harmonic function made: (1) of the
sum of the three harmonics found in the sunspot record modulated by
their own millennial beat harmonic; (2) of the astronomically deduced
phases of the tidal harmonics generated by Jupiter and Saturn; and
(3) on the sunspot number record from 1750 to 2010. The Sun is then
assumed to process the input harmonic function via numerous internal
mechanisms that are very reasonably non-linear. See the Appendix for
a summary of the equations of the model. 

The final output functions, that is what is measured and analyzed
such as the $^{14}C$ and $^{10}Be$ records, are non-linear modification
of the input three-frequency harmonic function. As well know by simple
algebraic relation, non-linear processing of a harmonic function generates
additional frequencies associated to the mutual beats and sub-harmonics
of the input generating frequencies. For example, if an input function
made of two harmonics as $f(t)=\cos\left(2\pi f_{1}t\right)+\cos\left(2\pi f_{2}t\right)$
is non-linearly processed into the function $g(t)=\left(1+f(t)\right)^{2}$,
the latter is characterized by 6 harmonics: $f_{1}$, $f_{2}$, $2f_{1}$,
$2f_{2}$, $|f_{1}+f_{2}|$ and $|f_{1}-f_{2}|$. 

More complex non-linearizations of an input harmonic function would
produce more harmonics but these are always related to the same mutual
beats and sub-harmonics of the generating frequencies. Thus, any generic
non-linearization of the input three-frequency model would generate
a similar set of theoretical expected frequencies. Therefore, there
is no need to know the exact physical functions of the internal solar
processing mechanisms to determine the expected frequencies that can
emerge when a non-linear system is forced by a harmonic forcing. 

Because the sunspot number record presents minima that converge to
about zero, which indicates the existence of some activation threshold
and other non-linearities in the solar internal mechanisms, \citet{Scafetta2012c}
schematically represented the process by generating an output signal
where the negative values of the input original harmonic sequence
are set to zero (see Figure 2E and the Appendix). 

\citet{Scafetta2012c}'s three-frequency non-linear solar model was
proposed and must be interpreted as a first approximation or schematization
of solar output variability. Other harmonics generated by the other
planets are surely present and may add some modulation, but they were
ignored because we would like to interpret the periods at 103, 115,
130 and 150 years. The three-frequency solar model produces at least
four main theoretical beat frequencies including those at about 61
years (beat between 9.93 and 11.86 year periods), 114.8 years (beat
between 9.93 and 10.87 year period), 130.2 years (beat between 10.87
and 11.86 year period) and about 983 years (combined beat among the
three harmonics) \citep{Scafetta2012c}. Similar harmonics are among
those typically observed in long solar records \citep[cf. ][]{Ogurtsov}.

\citet{Scafetta2012c}'s three-frequency non-linear solar model curve
is shown in Figure 2B against two nucleotide ($^{10}Be$  and $^{14}C$)
reconstructions of solar activity \citep{Bard,Steinhilber}. The schematic
signal can be further smoothed and rescaled to actual solar activity
records, e.g. the ACRIM total solar irradiance record \citep{ScafettaW2014},
to represent an ideal solar activity function generated by the chosen
three harmonics: see Figure 2C. Both Figures 2B and 2C show a clear
100-150 year synchronized modulation both in the nucleotide solar
proxy models and in the three-frequency solar model. The 100-150 year
scale reproduced by the model regulates and is in good phase with
the grand solar oscillations known as the Wolf, Sp\"orer, Mauder
and Dalton grand solar minima. \citet{Scafetta2012c} also showed
that the millennial cycle produced by the three-frequency non-linear
solar model is in good phase with the millennial oscillation observed
in typical solar and climate records throughout the Holocene for about
10,000 years used in \citet{Bond}. Figure 2F compares the millennial
oscillation predicted by the three-frequency non-linear solar model
(blue curve) versus the TSI proxy model by \citet{Steinhilber} (red
curve), and a good phase correlation can be discerned at this time
scale throughout the Holocene. 

Figure 2D shows the periodogram of the smooth three-frequency solar
model record and it shows main spectral peaks at 57, 61, 65, 103,
115, 130, 150 years, in addition to other minor harmonics and the
original three generating frequencies at 9.93, 10.87, 11.86 years.
The frequencies highlighted in Figure 2D are independent of the smooth
algorithm and are found in the record depicted in Figure 2B. Aliasing
artifacts noted by \citet{Poluianov} referring to the model proposed
by \citet{Abreu} are not present in my analysis because the three-frequency
model is constructed with three decadal harmonics and the periodogram
is calculated with a record sampled every 6 months for 10,000 years.

In conclusion, both \citet{Cauquoin} and \citet{Steinhilber} $^{10}Be$
and $^{14}C$ solar proxy models present 4 common spectral peaks at
about 103, 115, 130 and 150 years (confidence level $\sim95\%$ or
larger). Two of these frequencies appear to be predicted by \citet{Abreu}
planetary model, although with some doubt \citep{Poluianov}. However,
all four frequencies are predicted by the three-frequency non-linear
solar-planetary model proposed by \citet{Scafetta2012c} that combines
the major Jupiter and Saturn's planetary decadal harmonics with the
Schwabe 11-year solar cycle and further requires a non-linear processing,
whose existence is very reasonable given the fact that both solar
and geophysical systems are non-linear. 

\citet{Abreu}'s model does not reproduce these cycles because the
four oscillations at 103, 115, 130 and 150 year emerge from the modulation
of the 10.87 year central sunspot cycle, which is likely generated
by the solar dynamo, by the two side Jupiter and Saturn planetary
tidal harmonics shown in Figure 2A plus the nonlinear solar response
to external harmonic forcing. \citet{Abreu}'s model does not reproduce
the 11-year solar cycle and, therefore, it cannot reproduce the four
secular oscillations found in the solar proxy records depicted in
Figure 1. 

In conclusion, the analysis contradicts \citet{Cauquoin}'s conclusion.
I have demonstrated that common evidences do exist for a planetary
influence on solar activity 330,000 years ago and during the Holocene
once that the appropriate planetary-solar model, that is the three-frequency
non-linear solar model \citep[see also the Appendix]{Scafetta2012c,Scafetta2012d},
is used. Other harmonics found in \citet{Abreu} can be interpreted
in alternative ways. For example, the $\sim87$-year harmonic seems
to be related to Jupiter, Uranus and Neptune \citep[cf: ][]{Scafettaw2013a},
and the $\sim207$-year harmonic may be a beat between the $\sim61$-year
harmonic and $\sim87$-year harmonic that, through non-linear processing,
produces a $\sim205$-year cycle.

\section{The decadal and secular scale spectral coherence between the solar
system and the Earth's global surface temperature }

\citet{Scafetta2010,Scafetta2012a,Scafetta2012b,Scafetta2013a,Scafetta2013b}
found that the climate system since 1850 presents evidences of multiple
astronomical harmonics at the periods of about 5.2 year, 5.93 year,
6.62 year, 7.42 year, 9.1 year (main solar-lunar tide cycle), 10.4
year (related to the 9.93-10.87-11.86 year solar cycle harmonics),
13.8 year, $\sim$20 year, $\sim$30 year and $\sim$61 year \citep[cf.: ][]{Scafetta2010,Scafetta2012a,Scafetta2012c,Scafetta2012d,Scafetta2014,Wang2012}.
This property was found by direct comparison between the temperature
and astronomical spectra and by taking into account evidences for
paleoclimatic temperature oscillations and the basic harmonics known
from astronomy. For the benefit of the reader Figure 12 in the Appendix
reproduces figure 6B and 9A of \citet{Scafetta2010} showing {[}A{]}
the $\chi^{2}$ spectral coherence test and {[}B{]} the direct comparison
between the MEM curve of several climatic records and the astronomical,
solar and lunar harmonics (green bars). The spectral coherence between
the two systems is quite evident at multiple periods. 

Using time frequency analysis based on $L=60$ year moving windows
and magnitude squared coherence analysis based on $L=20$ and 30 year
moving windows \citet{Holm} questioned my results by claiming that
he could find only a frequency \textit{``line in the 15-20 year range
which varies with time as well as one around 9 years which comes and
goes and varies in frequency''}: see Figure 3. \citet{Holm} interpreted
his results as contradicting Scafetta's hypothesis of a possible astronomical
origin of the solar/temperature oscillations because Scafetta's astronomical
hypothesis would imply multiple time quasi-invariant temperature spectral
lines at specific astronomical frequencies. 

It is evident, however, that results are not independent of the mathematical
methodology adopted for the analysis. Specific physical properties
can be well highlighted with an analysis methodology but not equally
well with a different one. Thus, the mere fact that Holm could not
find Scafetta's results by using a different analysis methodology
is not a sufficient argument to dismiss Scafetta's claims.

Below I will clarify several physical and mathematical issues and
demonstrate that Holm's conclusion is flawed because his spectral
analysis does not have a sufficient spectral resolution and, therefore,
it was highly inadequate to properly identify the harmonics of interest
and determine the expected spectral coherence patterns found in \citet{Scafetta2010}.

\subsection{Summary of Holm's analysis}

\begin{figure}[!t]

\centering{}\includegraphics[width=0.8\textwidth]{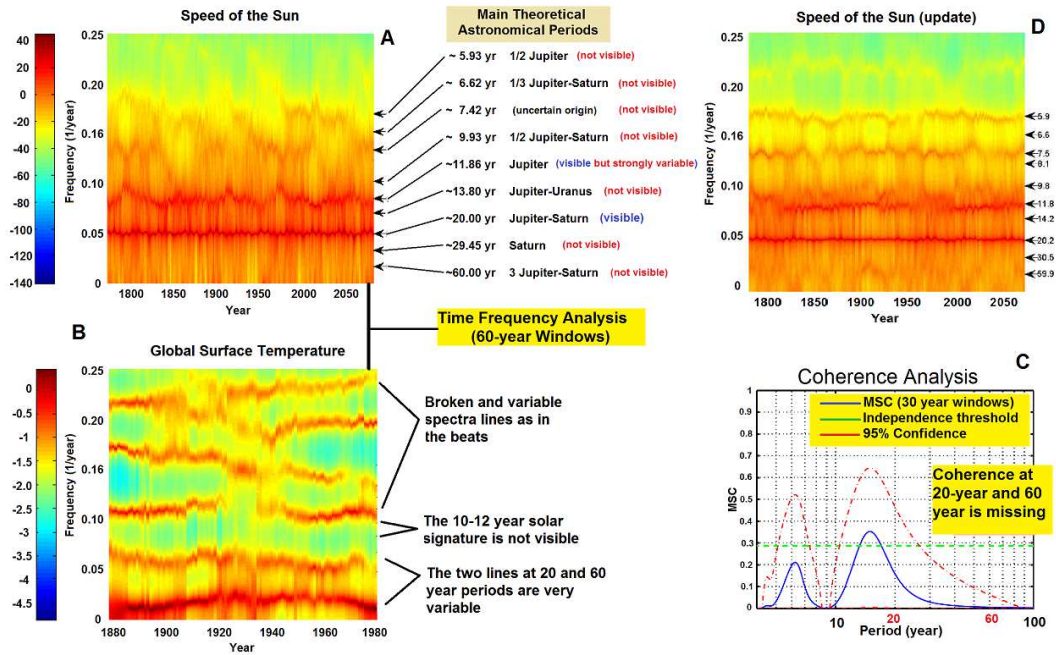}\protect\caption{{[}A,B,D{]} Time frequency analysis with Maximum Entropy Method (MEM)
power spectrum with an order M=42\% of the window length. {[}A{]}
Speed of the Sun relative to the barycenter of the solar system (using
a windows length $L<60$ years, probably $L=40$ years). {[}B{]} Earth's
global surface temperature. {[}C{]} Magnitude squared coherence using
$L=30$ year windows between the two records: peaks are at 6.3 and
14.9 years. The embedded table lists the main (orbital and synodic)
harmonics theoretically expected in the astronomical record. Most
of them are not visible in the Holm's graph, or their signature is
strongly intermittent and variable. Compare with Figures 8 and 10
where the analysis is repeated with $L=110$ years. The figures are
adapted from figures 2, 3 and 5 in \citet{Holm}. Figure D shows a
corrigendum of Figure 3A based on a windows length $L=60$ years \citep{Holmcc},
but still many astronomical harmonics are missing and/or strongly
variable.}
\end{figure}

Figure 3A shows \citet[figure 3]{Holm}'s time frequency analysis
of the speed of the Sun relative to the barycenter of the solar system.
This figure already suggests that Holm's time frequency analysis is
inadequate to identify the harmonics of interest. In fact, among the
main theoretical astronomical harmonics (reported in Figure 3 as well)
only the strong $\sim20$-year synodic period between Jupiter and
Saturn is clearly visible. Jupiter's 11.86-year orbital period is
weakly visible and very unstable: it appears to vary between 10 and
14 years. All other expected astronomical harmonics are not visible
or are strongly intermittent and variable. 

Holm himself seemed to be partially aware of the problem and in April,
2014 he published on arXiv a corrigendum \citep{Holmcc} with an updated
figure 3 herein reported in Figure 3D. This corrigendum was required
because \citet{Holm} claimed that his figure 3 (depicted herein in
Figure 3A) used a time frequency analysis with window length $L=60$
years, while indeed the graph was based on a shorter window length
($L=40$ years). His updated figure (depicted herein in Figure 3D)
shows the analysis with window length $L=60$ years. Now, the astronomical
frequencies at periods of 5.9 and 7.4 years appear, and the 11.86
year Jupiter orbital period becomes more stable. However, many expected
astronomical harmonics are still missing, and those visible are strongly
variable with the exception of the dominant 20-year Jupiter-Saturn
synodic oscillation. Yet, the astronomical harmonics are quite stable
in reality. 

The missing and/or variable spectral line problem is due to the adoption
of a too short window length $L$ in the spectral analysis that yields
an inadequate spectral resolution required to solve close harmonics.
Indeed, comparing Figure 3A (using $L=40$ years) and its updated
Figure 3D (using $L=60$ years) clearly suggests that by simply increasing
the time frequency window length, new spectral lines appear and others
become more stable. Holm's also used a too small MEM order M. He used
$M=42\%$ of the data length, while it would have been better to use
the maximum available value, that is $M=50\%$ of the data length,
because of the very short windows he used. The same problem also corrupts
Holm's time frequency analysis for the temperature record shown in
Figure 3B and his spectral coherence analysis.

\subsection{Understanding the first and second law of Kepler, and the data patterns}

Some of the shortcomings of Holm's analysis are self-evident from
his own figures once basic astronomy is taken into account. The purpose
of \citet{Scafetta2010}'s analysis was to identify the astronomical
frequencies and test whether they could be found in the climate records.
However, in his coherence figures 4 and 5, which were made using windows
with lengths of $L=20$ and 30 years respectively, \citet{Holm} could
not find \citet[figures 6, 9 and 10 ]{Scafetta2010}'s spectral coherence
between the Earth's temperature and the astronomical record at the
$\sim20$ and $\sim60$ year periods, as well as at other frequencies.
Compare versus \citet{Scafetta2010} results briefly summarized in
Figure 12 in the Appendix.

Missing the coherence at the 20-year period is paradoxical because
in his own time frequency analysis Holm could find a quasi 20-year
oscillation in both the temperature and astronomical record: see Figure
3. On the contrary, the missing coherence at the 60-year period was
due to the fact that while Holm could find a quasi 60-year oscillation
in the temperature record (see Figure 3B), both in his original and
in his updated figure (see Figures 3A and 3D) he could not find a
distinct 60-year oscillation in the astronomical record. In Section
3.9 I will show that this failure is due to the $L=20$ and 30 year
windows used in his coherence analysis, which are even shorter than
the $L=60$ year window used in his time frequency analysis.

Note that the same $L=60$ year time frequency technique could find
the 60-year oscillation in the temperature record but not equally
well in the astronomical record because the sensitivity of spectral
analysis also depends on the actual spectral power of the harmonics.
Small harmonics are harder to highlight if their period is close to
the length of the data and other close harmonics have a larger power.
In the temperature record the 60-year oscillation is the dominant
cycle as shown in Section 3.6, while in the chosen astronomical record
the 60-year oscillation is a secondary cycle and is significantly
smaller than the 20-year cycle. Yet, a 60-year cycle is clearly visible
in the JPL's HORIZONS ephemeris of the Sun's barycentric speed after
some moving average filtering to remove the 20-year oscillation: see
Figure 4C \citep[cf. ][]{Scafetta2010,Scafetta2014}. 

The existence of a 60-year astronomical oscillation cannot be questioned
in the barycentric speed of the sun because it derives directly from
the 5/2 resonance between 5 Jupiter's revolutions (orbital period
= 11.86 year) and 2 Saturn's revolutions (orbital period = 29.46 year).
The quasi 60-year oscillation emerges in the barycentric movement
of the Sun because according the first law of Kepler the orbits of
Jupiter and Saturn are elliptical; and according the second law of
Kepler their speeds change in function of the planetary coordinates
and determine the dynamics of the Sun's barycentric wobbling. 

Jupiter-Saturn conjunctions occur every 19.86 years and two consecutive
ones are separated by about $242.8{}^{o}$: see Figure 13 in the Appendix.
Thus, every three consecutive conjunctions, that is a \textit{trigon}
or about 60 years\textit{,} a Jupiter-Saturn conjunction occurs approximately
at the same position of the sky and, therefore, their combined position
and velocity repeats not only every $\sim20$ year but also every
$\sim60$ years. The trigon slightly rotates every 800-1000 years
giving origin to an additional quasi millennial oscillation: see the
detailed discussion in \citet{Scafetta2012a}, \href{http://en.wikipedia.org/wiki/Great_conjunction}{http://en.wikipedia.org/wiki/Great\_{}conjunction}.
The 20, 60 and 800-1000 year Jupiter-Saturn conjunction oscillations
were even well-known since antiquity and linked to historical events
that today would be linked to climate changes \citep[cf. ][]{Masar,Kepler}:
for example the 60-year oscillation was chosen as the basis for the
traditional Indian and Chinese calendars, where it is known as the
Brihaspati (that means \textit{Jupiter}) cycle, and likely linked
to the monsoon 60-year cycles \citep[cf. ][]{Agnihotri,Iyengar,Temple}. 

From the above, it is evident that Holm's failure of clearly detecting
the 60-year oscillation in the Sun's movement could only mean that
Holm's analysis was inadequate to detect accurately the elliptical
shape and the dynamics of the orbits of Jupiter and Saturn in the
Sun's speed record. Therefore, Holm's methodologies had to fail to
find the spectral coherence between astronomical and temperature frequencies
because they could not find the frequencies of interest in the first
place. 

In any case, it is also important to highlight that if the planetary
record is substituted with the tidal function, the quasi 60-year oscillation
emerges as a dominant beat cycle between the 9.93-year Jupiter-Saturn
spring tidal harmonic and the 11.86-year Jupiter orbital tidal harmonic,
which non-linear mechanisms transform into an explicit harmonic: see
Figure 2 \citep{Scafetta2012c,Scafetta2012d}. Therefore, there are
several ways to interpret the quasi 60-year oscillation using planetary
harmonics.

\subsection{The complex planetary synchronization structure of the solar system}

To test whether the climate presents a signature of astronomical frequencies
the first step is to determine them. Studying their physical and mathematical
properties is necessary to select the appropriate analysis methodology
to be used for the task.

It is important to understand that \citet{Scafetta2010} used the
speed of the Sun relative to the barycenter of the solar system only
and exclusively as a proxy for easily determining the main frequencies
of the gravitational oscillations of the heliosphere. Using different
planetary functions is expected to yield similar frequencies because
any function of the planetary orbits present the same set of frequencies
by algebraic construction. However, each individual function stresses
them differently by showing alternative spectral amplitudes.

On the contrary, to deduce the physical amplitudes all physical mechanisms
linking the astronomical to the climate records need to be identified.
These also include the internal resonance frequencies of the Sun and
of the climate system, and their numerous feedback mechanisms. To
develop a full physical model that explains in details both solar
dynamics and the Earth' climate using planetary forces is far beyond
the purpose of the present paper and of the present scientific knowledge.

The frequencies of interest can be theoretically deduced from astronomical
considerations alone by taking into account the orbital periods of
the planets, their combinations and various sub-harmonics. However,
numerous frequencies are expected, and it is simpler to calculate
them by analyzing the power spectrum of a planetary proxy record such
as those describing the Sun's barycentric movement.

\citet{Scafetta2014} used the JPL\textquoteright s HORIZONS Ephemeris
system to calculate the wobbling and the speed of the Sun from 12
December 8002 BC to 24 April 9001 AD (100-day steps): see Figure 4.
Perhaps, the speed of the wobbling Sun can have an electromagnetic
interaction meaning. On the contrary, for gravitational interactions
the tidal function should be used and an example was discussed in
the previous section that stresses differently some harmonics, see
Figure 2 \citep{Scafetta2012a,Scafetta2012c,Scafetta2012d,ScafettaW2013,Scafettaw2013a}.
However, here the issue is to determine the basic frequencies of the
solar system.

\begin{figure}[!t]
\centering{}\includegraphics[width=0.8\textwidth]{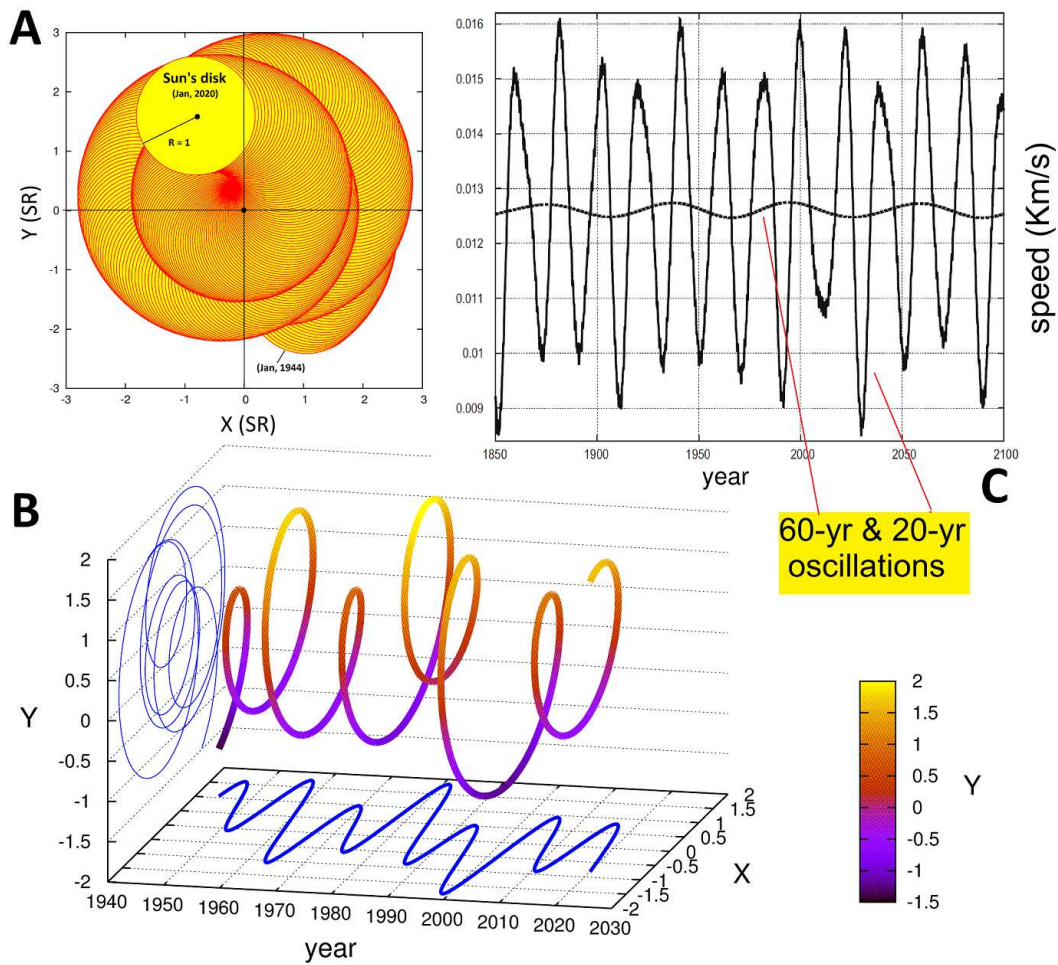}\protect\caption{The wobbling of the Sun relative to the center of mass of the solar
system. (A) Monthly scale movement of the Sun from 1944 to 2020 as
seen from the z axis perpendicular to the ecliptic. The Sun is represented
by a moving yellow disk with a red circumference. (B) The trajectory
of the center of the Sun from 1944 to 2020. (C) The speed of the Sun
from 1850 to 2100. Note the evident 20-year and 60-year oscillations:
the latter is highlighted by the smooth dash curve. The Sun\textquoteright s
coordinates are estimated using the Jet Propulsion Lab\textquoteright s
(JPL) HORIZONS Ephemeris system. The coordinates are expressed in
solar radius (SR) units. The figures are adapted from \citet{Scafetta2010,Scafetta2014}.}
\end{figure}

Power spectrum of the Sun's speed was evaluated using the Multi Taper
Method (MTM) \citep{Ghil2002}: see Figure 5. The sub-annual spectrum
of this record is shown and studied in \citet{ScafettaW2013}, where
it was found coherent with several total solar irradiance high frequency
harmonics. 

Main spectral peaks at about 5.93, 6.63, 7.42, 9.93, 11.86, 13.8,
20, 60.9 years are clearly observed: see the table depicted in Figure
3 for their physical attribution. Note that some spectral peaks are
common with the independent three-frequency solar model based on tidal
cycles and the sunspot record (Figure 2D) such as the triplet at 57,
60.9 and 65 year period that emerges from an asymmetry of the 60-year
\textit{trigon} cycle. 

A major physical result of the analysis is that, as shown in Figure
5, from the annual to the secular scale, the natural frequencies of
the solar system (shown by the peaks in the red MTM curve) are approximately
reproduced by the following simple empirical harmonic formula \citep{Jakubcova,Scafetta2014}:

\begin{equation}
f_{i}=\frac{1}{P_{i}}\approx\frac{i}{178.4}\qquad yr^{-1},\qquad i=1,2,3,\ldots\label{eq:1-1}
\end{equation}
Eq. \ref{eq:1-1} suggests that the solar system is almost locked
in a synchronization pattern made of a specific harmonic series. The
finding can be considered compatible with Kepler\textquoteright s
vision of a \textit{cosmographic mystery} and of the \textit{harmonices
mundi} (the harmony of the world), known since Pythagoras of Samos
as \textit{musica universalis} (universal music or music of the spheres).
Because the entire solar system appears to be subtly self-synchronized
in a harmonic sequence, the frequencies of its coupled gravitational
structure could be expected to influence also solar dynamics and the
Earth's climate.

\begin{figure}[!t]
\centering{}\includegraphics[angle=-90,width=0.8\textwidth]{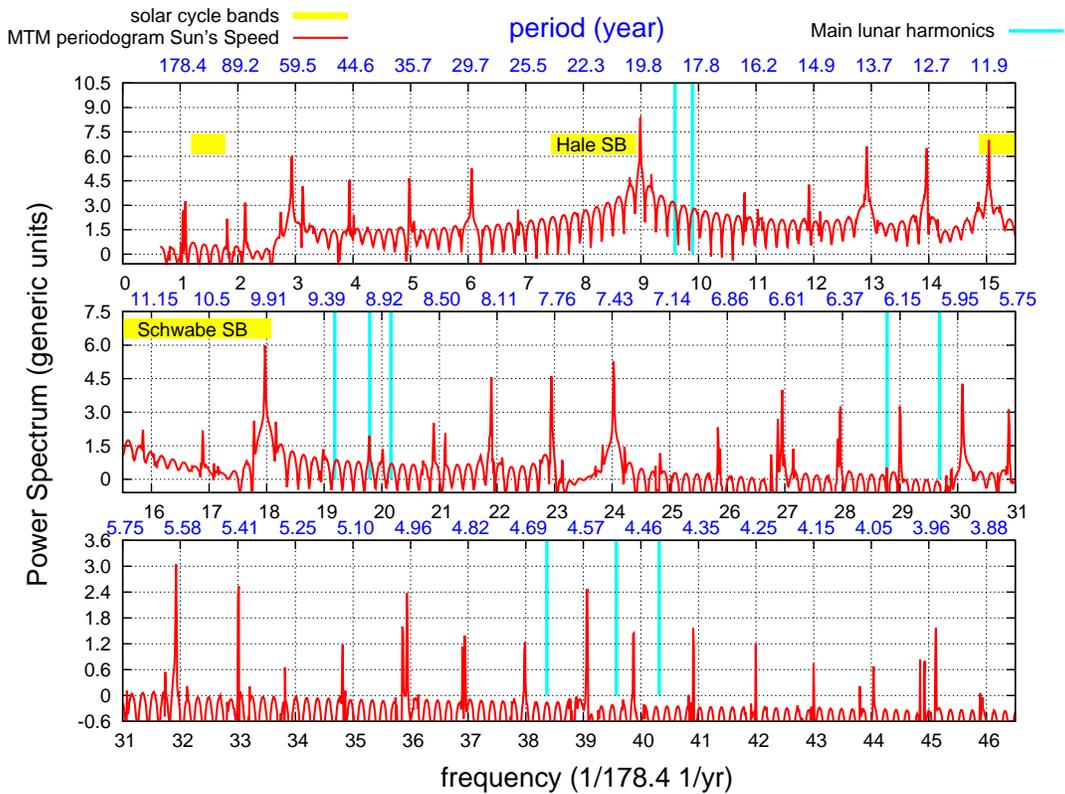}\protect\caption{MTM periodogram (red) of the speed of the Sun relative to the center
of mass of the solar system from Dec 12 8002 B.C. to 24 Apr 9001 A.D.
All spectral peaks are highly significant because the astronomical
record has likely of 6-7 digit precision and the peaks are sharp and
can be recognized from astronomical consideration alone because correspond
to the orbital periods, the synodic periods and their harmonics. The
cyan lines correspond to theoretical lunar tidal harmonics that can
influence the climate: the 18.6, 18.03, 8.85 year cycles and their
harmonics. The yellow areas correspond to the Schwabe 11-year solar
cycle band, to the Hale 22-year solar cycle band and the 100-150 year
solar band discussed in Figure 2. }
\end{figure}

In fact, as discovered by Huygens in the 17th century, in absence
of any significant noise or friction (in the space there is none)
coupled oscillators (e.g. the Sun, the planets and the moons of the
solar system) can mutually synchronize also if the strength of the
physical interactions is quite weak \citep{Pikovsky}. \citet{Scafetta2014}
provided a general introduction on the complex planetary synchronization
structure of the solar system. There a reader can find a review of
the relevant research. This includes topics such as: Copernicus and
Kepler\textquoteright s vision of a cosmographic mystery revealing
a \textit{\textquotedblleft marvelous proportion of the celestial
spheres\textquotedblright{}} referring to the \textit{\textquotedblleft number,
magnitude, and periodic motions of the heavens\textquotedblright{}}
\citep{Copernicus,Kepler1596}; the planetary rhythm of the Titius\textendash Bode
rule \citep{Titius}; the asteroid belt \textit{mirror} symmetry rule
\citep{Geddes}; the matrix of planetary resonances \citep{Molchanov1968,Molchanov1969a,Molchanov1969b};
the 2/3 synchronization of Venus axis rotation period with the Earth's
orbital period so that at every inferior Venus-Earth conjunction the
same side of Venus faces Earth \citep{Goldreich}; the existence of
a mathematical relation linking planetary orbital parameters \citep{Tattersall};
the existence of several other celestial commensurabilities such as
the quasi synchronization between the 27.3-year lunar orbital period
with the 27.3-year Carrington solar rotation cycle as seen from Earth;
and many others. 

Figure 5 also shows in cyan the major theoretical lunar tidal harmonics
deduced from the 18.6-year lunar nodal cycle, the 18.03-year Saros
cycle, the 8.85 apside rotation cycle and their second, third and
forth harmonics. The yellow areas represent the Schwabe 11-year solar
cycle band, the Hale 22-year solar magnetic cycle band and the 100-150
year solar band discussed in Figure 2.

The frequencies shown in Figure 5 represent the theoretical astronomical
harmonics that could be reasonable expected to influence the climate.

\subsection{Understanding the spectral resolution of the time frequency analysis}

The astronomical sequence analyzed in Figure 5 is about 17,000 years
long and its spectral peaks are sharp and many of them can be easily
recognized from simple astronomical considerations alone. However,
short sequences theoretically made of the same harmonics are difficult
to analyze because close harmonics generate complex low frequency
beats that need to be solved by the spectral analysis to separate
them. In fact, the closer the frequencies, the longer the segment
must be to conclude that more than one frequency is present within
a specific spectral band. If the segment is too short, two close waveforms
will look like a single sine wave and the spectral technique fails
to separate them. In brief, as well known, the length of the signal
limits the frequency resolution of the spectral analysis. Let us clarify
the issue using mathematics.

If $T$ is the acquisition time of a sequence, $N$ is the number
of acquired samples and $f_{s}$ is the sampling frequency, the frequency
resolution of its Fourier analysis is defined as (\citeauthor{FFT}):

\begin{equation}
\nabla f=\frac{1}{T}=\frac{f_{s}}{N}.
\end{equation}
A frequency $f$ associated with a spectral peak has an uncertainty
of $\pm\text{\textonehalf}\nabla f$ and the correspondent period
$P$ is:

\begin{equation}
P=\frac{1}{f\mp\text{\textonehalf}\nabla f}=\frac{1}{f}\pm\frac{\nabla f}{2f^{2}}=\frac{1}{f}\pm\frac{P^{2}}{2T}.\label{eq:error}
\end{equation}
To well separate two frequencies $f_{1}=1/P_{1}$and $f_{2}=1/P_{2}$,
the following condition must be fulfilled

\begin{equation}
f_{12}=|f_{1}-f_{2}|\gtrapprox\nabla f,\label{eq:4-1}
\end{equation}
where, as well known, $f_{12}=|f_{1}-f_{2}|$ is the beat frequency
between the two frequencies. Consequently to well separate two close
frequencies using Fourier based spectrum analysis the acquisition
time $T$ must be about or larger than the beat period, $P_{12}=1/f_{12}$,
between the two contiguous harmonics, that is: 

\begin{equation}
T=\frac{1}{\nabla f}\gtrapprox\frac{1}{|f_{1}-f_{2}|}=P_{12}.\label{eq:beat}
\end{equation}

The Maximum Entropy Method (MEM) produces sharper spectral peaks with
smaller error bars, but MEM can also produce spurious spectral peaks.
For this reason it is conventional to validated the MEM spectral peaks
using Fourier based periodograms \citep{Press}, and then use MEM
to sharp the results, as I have often done \citep[e.g.: ][]{Scafetta2010,Scafetta2012b,Scafetta2013b}.
So, for safety, the condition of Eq. \ref{eq:beat} must be used also
when MEM is applied.

Time-frequency analysis divides a sequence in moving window segments
of length $L$ and evaluates the spectra of these segments to study
how the spectral energy evolves in time. Eq. \ref{eq:beat} implies
that, when time-frequency analysis is used to determine whether a
sequence contains a specific set of time-invariant spectral lines,
it is necessary to chose its moving-window length $L$ to be about
or longer than the theoretical beat periods among the proposed harmonic
constituents of the signal. If $L\ll P_{beat}$ time-frequency analysis
only highlights the interference dynamics of the beats among the constituent
harmonics producing variable patterns such as those visible in Figure
3. 

\begin{figure}[!t]
\centering{}\includegraphics[width=0.8\textwidth]{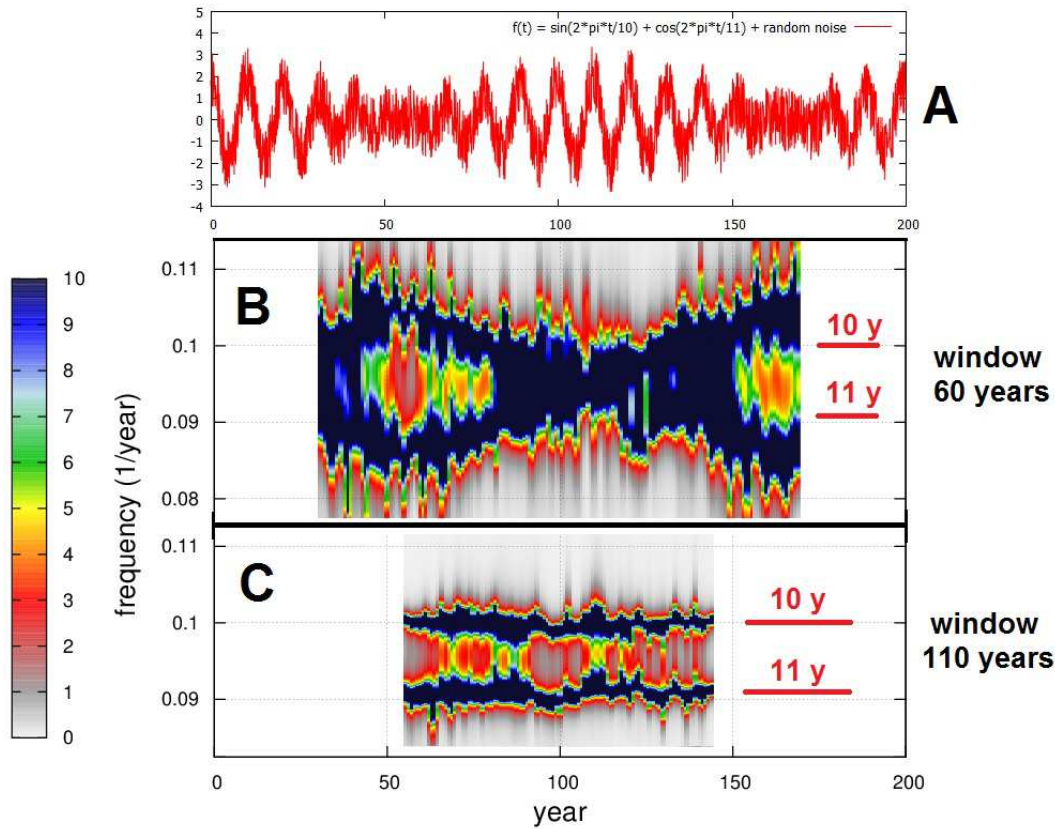}\protect\caption{{[}A{]} Artificial sequence made of two harmonics with periods of
10 and 11 years, plus random noise: the record shows a clear periodic
beat with a 110-year period. {[}B{]} Time-frequency analysis of the
signal using MEM and $L=60$ year moving windows. A variable pattern
is observed. {[}C{]} Time-frequency analysis of the signal using MEM
and $L=110$ year moving windows, which is the theoretical beat period.
The two constituent harmonics are highlighted by the analysis. The
MEM order is M=50\% of the available points ($M=N/2=360$ and $M=660$,
respectively). The panel colors are in generic units from a minimum
power indicated as ``0'' to a maximum indicated as ``10''.}
\end{figure}

Let us explain this result with a simple exercise that also serves
to illustrate the inadequacy of Holm's 60-year and shorter window
methodologies. Figure 6A shows an artificial sequence of 200 years
made of two harmonics with periods of 10 and 11 years plus random
noise. Figure 6B shows its MEM time-frequency analysis using 60-year
windows as chosen in \citet{Holm}. It is evident that with $L=60$
years the technique fails to emphasize the presence of the two constituent
time-invariant harmonics by producing a variable beating pattern\textit{.
}In fact, the frequency resolution associated to a 60-year window
is $\nabla f_{60}=1/60=0.0167$. This value is significantly larger
than the minimum spectral resolution required to separate the two
harmonics, which is $\nabla f_{10,11}=1/10-1/11=1/110=0.0091$. Figure
6C shows the time-frequency analysis of the signal using $L=110$
year windows, which corresponds to the beat period. Now the time-frequency
analysis clearly reveals the two expected time-invariant spectral
constituent lines at the periods of 10 and 11 years.

\subsection{The optimal spectral window length}

In choosing the most appropriate window for his time frequency analysis
\citet{Holm} simply argued: \textit{``The length of the data window
must be chosen and the longer the window, the better the frequency
resolution. On the other hand, the shorter the window, the better
the ability to track time variations in the data. For the examples
shown here, a window length of 60 years was found to be a reasonable
compromise.''} However, it appears that Holm's \textit{``reasonable
compromise''} was based only on a qualitative personal impression.
In fact, it was not based on any mathematical or physical property
of the data. 

It is true that using windows as short as possible increases the number
of independent spectra that can be compared. However, Holm ignored
that, to be useful, the frequency resolution of the methodology must
be appropriate to separate the harmonics of interest. Thus, the window
length could be chosen to be small but not smaller than the largest
beat period among the contiguous frequencies that one is interested
to identify. 

According Eq. \ref{eq:1-1} and Figure 5 the theoretical expected
astronomical harmonics in the frequency range of interest are approximately
sub-harmonics of a 178.4-year oscillation \citep[cf.][]{Jakubcova,Scafetta2014}.
This means that the minimal spectral resolution $\nabla f$ required
to identify them in both the astronomical and temperature record is

\begin{equation}
\nabla f=\frac{1}{L}\lessapprox\left|f_{i+1}-f_{i}\right|=\frac{i+1}{178.4}-\frac{i}{178.4}=\frac{1}{178.4}=0.0056\quad yr^{-1}.\label{eq:res}
\end{equation}
Eq. \ref{eq:res} indicates that the spectral window length should
be chosen to be $L\gtrapprox178.4$ years. This length is not only
far larger than Holm's 20-to-60 year spectral windows, but it is also
larger than the available 164-year temperature record. Thus, using
a single window equal to the length of the entire temperature record,
as done in \citet{Scafetta2010} and below in Table 1, is right now
the most appropriate way to study the spectral coherence between astronomical
and climatic harmonics. Because 164 years is just slightly shorter
than 174 years, MEM could still be reliable to get all main expected
harmonics with sufficient sharpness using a single 164-year window
under the condition of using the maximum MEM order (M=50\% of the
available data), as done in \citet{Scafetta2010}.

\subsection{Estimation of the expected main beat periods in a 164-year record}

To highlight at least the major visible harmonics using time frequency
analysis, one needs to study the major beat periods among those clearly
visible in a 164-year short record. 

\citet{Holm} used the HadCRUT3 global surface temperature record
\citep{Brohan2006}. Herein I re-analyze this record from Jan/1850
to Dec/2013. It is made of 1968 monthly data or 164 annual data. Figure
7A shows the annual temperature record from 1850 to 2013 ($N=164$
years) with its error bars. Figure 7B shows its power spectrum functions
calculated with MEM (using $M=N/2=82$) and with MTM \citep{Ghil2002}.

I used both techniques because, as \citet[pp. 574]{Press} wrote,
\textit{\textquotedbl{}Some experts recommend the use of this algorithm
in conjunction with more conservative methods, like periodogram, to
help choose the correct model order, and to avoid getting too fooled
by spurious spectral features.\textquotedbl{}} Essentially, first
the periodogram is used to find the spectral peaks with their statistical
confidence, then MEM is used to sharp the results. MEM spectral peaks
that are not confirmed by the periodogram should be rejected, and
vice versa. 

Figure 7B does not show any MEM spurious peaks because each peak is
confirmed by a correspondent peak in the MTM periodogram. Thus, it
is safe to use the maximum MEM order, $M=N/2$, that adds sharpness
to well solve the low frequency range of the record that contains
both the long harmonics (e.g. the 20-year and 60-year cycle) and the
beat harmonics of the fast oscillations \citep[cf.][, supplement file]{Scafetta2012b}.

As Figure 7B shows, most temperature spectral peaks for 4-year and
larger periods have a 99\% statistical confidence relative to the
physical noise (gray area in Figure 7A) \citep{Ghil2002}. The temperature
error has an average standard deviation of $\sigma\approx0.06$ $^{o}C$.
The 95\% and 99\% spectral confidence levels were deduced from computer
generated random Gaussian noise with $\sigma\approx0.06$, and using
the SSA-MTM tool kit for spectral analysis \citep{Ghil2002}; they
are approximately given by the equation $8.5\sigma^{2}/\pi$ (95\%)
and $14.6\sigma^{2}/\pi$ (99\%), respectively, where $1/\pi$ is
the MTM spectral median for normal Gaussian noise. The MTM spectral
peaks have a statistical error given by Eq. \ref{eq:error}. For example,
for the 60-year cycle the error is $\pm11$ years. The MEM and MTM
specral peaks are consistent with each other within the MTM spectral
errors.

\begin{figure}[!t]
\begin{centering}
\includegraphics[width=0.8\columnwidth]{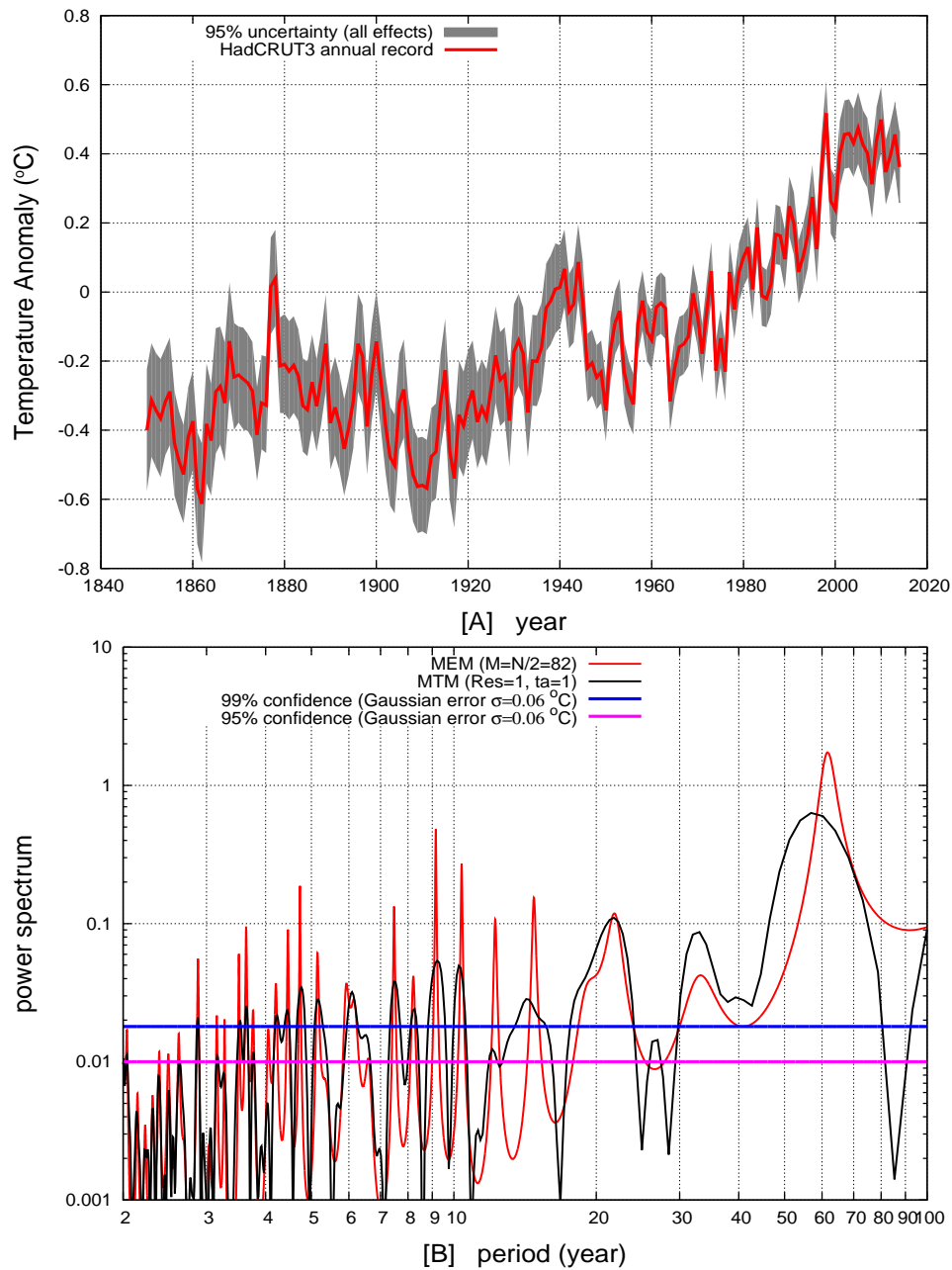}\protect\caption{{[}A{]} Annually solved HadCRUT3 record \citep{Brohan2006} from 1850
to 2013 (N=164 years). {[}B{]} Power spectrum functions calculated
with the MEM method (using M=N/2=82) and the MTM periodogram \citep[SSA-MTM Toolkit]{Ghil2002}.
The MTM power spectrum statistical confidence levels were deduced
using computer generated Gaussian random sequences simulating the
combined effects of all known physical temperature uncertainties (annual
average has $\sigma\approx0.06$ $^{o}C$). The temperature uncertainties
vary in time and are shown in A by the gray area at the $2\sigma$
level (95\% confidence). (\protect\href{http://www.metoffice.gov.uk/hadobs/hadcrut3/diagnostics/time-series.html}{http://www.metoffice.gov.uk/hadobs/hadcrut3/diagnostics/time-series.html})}

\par\end{centering}

\end{figure}

Table 1 reports: (1) all main spectral peaks found in the temperature
record in the range from 4 to 100 year period; (2) the main astronomical
expected periods from Figures 3 and 5; (3) $\chi^{2}$ index between
the temperature and astronomical periods (values of $\chi^{2}\lessapprox1$
indicate spectral coherence); (4) the theoretical beat periods among
the contiguous astronomical frequencies. The $\chi^{2}$ index values
between the astronomical and temperature main frequencies reported
in Table 1 reveals that these frequencies are coherent to each other
within the spectral resolution of the analysis.

Table 1 reports that for most frequencies the astronomical-temperature
coherence is quite good, $\chi^{2}<1$, while for others is still
good but $\chi^{2}\approx1$. This may be explained by unresolved
beats. For example, the temperature shows a spectral peak around $14.5$-year
period that could be compared with the closest dominant 13.8-year
Jupiter-Uranus period. However, Figure 5 reveals that in addition
to the 13.8-year harmonic, the astronomical record also contains a
15-year harmonic (=half orbital period of Saturn). So, it is possible
that the 14.5-year temperature harmonic is a compromise between the
two close astronomical harmonics that the spectral analysis could
not well separate. In fact, the beat frequency between the two harmonics
is $f_{12}=|f_{13.8}-f_{15}|=|1/13.8-1/15|=1/172.5=0.0058$ $yr^{-1}$,
which is shorter than the the spectral resolution available for the
164-year temperature signal, $\nabla f=1/164=0.0061$ $yr^{-1}$,
Eq. \ref{eq:4-1}. Some spectral disruption in the temperature record
may also derive from the smooth anthropogenic and volcano signatures
that this record contains \citep[e.g.: ][and many others]{Scafetta2013b},
but this topic is not addressed here further. 

\begin{table}[!t]
\centering{}%
\begin{tabular}{|c|c|c|c|c|}
\hline 
Temperature & Sun's Speed & $\chi^{2}$ & close periods & beat period (y)\tabularnewline
$P_{tem}$ (year) & $P_{sun}$ (year) & $\frac{(P_{tem}-P_{sun})^{2}}{(\nabla P_{tem})^{2}}$ & year : year & \tabularnewline
\hline 
\hline 
$5.2\pm0.08$ & 5.12 & 1.00 & 5.12 : 5.93 & 37.5\tabularnewline
\hline 
$5.95\pm0.11$ & 5.93 & 0.04 & 5.93 : 6.63 & 56.2\tabularnewline
\hline 
$6.54\pm0.13$ & 6.63 & 0.48 & 6.63 : 7.42 & 62.3\tabularnewline
\hline 
$7.5\pm0.17$ & 7.42 & 0.22 & 7.42 : 8.14 & 83.9\tabularnewline
\hline 
$8.25\pm0.21$ & 8.14 & 0.27 & 8.14 : 8.85 & 102\tabularnewline
\hline 
$9.1\pm0.25$ & 8.85-9.30$^{*}$ & 0.00 & 9.1 : 9.93 & 109\tabularnewline
\hline 
$10.4\pm0.33$ & 9.93-11.86$^{**}$ & 0.18 & 11.86 : 13.8 & 84.4\tabularnewline
\hline 
\multirow{2}{*}{$14.5\pm0.7$} & 13.8 & 1 & 13.8 : 19.86 & 45.2\tabularnewline
\cline{2-5} 
 & 15.0 & 0.51 & 15.0 : 19.86 & 61.3\tabularnewline
\hline 
$20.7\pm1.2$ & 19.86 & 0.49 & 19.86 : 29.45 & 61\tabularnewline
\hline 
$32\pm3.3$ & 29.45 & 0.60 & 29.45 : 60.9 & 57\tabularnewline
\hline 
$61\pm11$ & 60.9 &  &  & \tabularnewline
\hline 
\end{tabular}\protect\caption{(1) Spectral peaks of the HadCRUT3 global surface temperature monthly
record. The 61-year period has been optimized by comparison with the
periodogram and direct filtering \citep{Scafetta2010,Scafetta2013b}.
(2) Main spectral peaks of the Sun's speed: from Figure 5. $^{*}$Range
of the soli-lunar tidal periods which are not found in the Sun's speed.
$^{**}$Range of the harmonic constituents that make the sunspot cycle,
see Figure 2A \citep{Scafetta2012c}. (3) $\chi^{2}$ index between
the temperature and astronomical periods: values $\chi^{2}\lessapprox1$
indicate spectral coherence. (4) Main contiguous astronomical beating
frequencies expected to regulate temperature oscillations. (5) Average
beat periods. The astronomical spectral resolution is $\nabla f=1/17,000=0.00006$
$yr^{-1}$, while the temperature spectral resolution is $\nabla f=1/164=0.0061$
$yr^{-1}$ and it is used for estimating the error of the reported
temperature periods using Eq. \ref{eq:error}.}
\end{table}

The minimal windows length to be used in a time-frequency analysis
should at least be larger than the beat periods listed in Table 1.
\citet{Scafetta2010,Scafetta2012a,Scafetta2012b,Scafetta2012c,Scafetta2013a,Scafetta2013b}
showed that the quasi 9.1-year frequency found in the temperature
record may be bounded by the 8.85 years (the lunar apsidal precession)
and the 9.3 years (sub-harmonic of the 18.6-year lunar nodal cycle)
\citep[cf.][]{Wang2012} while the 10.4-year frequency is a variable
cycle related to the Schwabe solar cycle that derives from a combination
of the 9.93-year Jupiter-Saturn spring tidal cycle and the 10.87-year
central sunspot number spectral period: the average between the two
harmonics is 10.4 years \citep[cf.: ][]{Scafetta2010,Scafetta2012c,Scafetta2014}.
Because the 9.93-year cycle is a harmonic constituent of the 10.4-year
cycle, it bounds it. The beat period between the 9.1-year and the
9.93-year harmonics is about 108.9 years and it is larger than the
other beat periods listed in Table 1. Therefore, to highlight the
astronomical-temperature coherence at least for the harmonics most
visible in a 164-year long sequence, a reasonable minimal window length
should be $L\approx110$ years, not the $L\approx60$ years used by
Holm.

\subsection{Time frequency analysis using 110-year windows}

Figure 8 compares the time-frequency analysis of the speed of the
Sun relative to the center of mass of the solar system (panel A) and
of the HadCRUT3 monthly temperature record (panel B) using $L=110$
year window and $M=N/2=660$ (50\% of the number of available monthly
samples). The figure demonstrates the presence in the temperature
panel of multiple time-invariant spectral lines at the major expected
astronomical harmonics. The main astronomical periods of interest
here are at about 5.2 year, 5.93 year (Jupiter 2f orbital harmonic),
6.62 year (Jupiter/Saturn 3f synodic harmonic), 7.42 year, 9.1 year
(main solar-lunar tide cycle), 10.4 year (related to the 9.93-10.87-11.86
year solar/Jupiter/Saturn harmonics), 13.8 year (Jupiter/Uranus synodic),
19.86 year (Jupiter/Saturn synodic), 29.95 year (Saturn orbital, which
is quite weak) and $\sim60$ year (Jupiter/Saturn triple conjunction
and beat tide) \citep[cf.: ][]{Scafetta2010,Scafetta2012a,Scafetta2012c,Scafetta2012d,Scafetta2014,Wang2012}.
Among the astronomical harmonics the temperature shows sufficiently
stable oscillations at about 7.44, 9.1, 13.8, 20 and 60 years. 

In the 5.93-6.63 range the temperature  shows two variable lines that
are, however, well bounded by the two major expected theoretical astronomical
oscillations. Note that the spectrogram of the entire temperature
record (Figure 7B) shows that these two minor harmonics are relatively
weak and quite diffused. Probably the used spectral methodology based
on 110-year windows was not sufficiently robust to separate well these
spectral lines. In fact, according the full spectrogram of the Sun's
speed (Figure 5A) the 5.93-6.63 year period range is occupied by 4
astronomical harmonics at 5.93, 6.15, 6.38 and 6.62 year. The 3f harmonic
of the 18.6 lunar nodal cycle ($18.6/3=6.2$ years) and of the Saros
cycle ($18.03/3=6.01$ year) are also expected to be present: see
Figure 5. Thus, the 5.93-6.63 year period range could be modulated
by five close astronomical harmonics that are beating and could generate
the variable pattern observed in Figure 8B. Windows far longer than
$L=110$ years would be required to solve this frequency rang: $L\gtrapprox170$
years if only the Sun's speed harmonics are considered, which can
increase up to $L\gtrapprox762$ years if the lunar 3f nodal harmonic
are also present. Therefore, the 5.93-6.63 year range cannot be solved
well with the current temperature record.

\begin{figure}[!t]
\centering{}\includegraphics[width=0.8\textwidth,height=0.4\textheight]{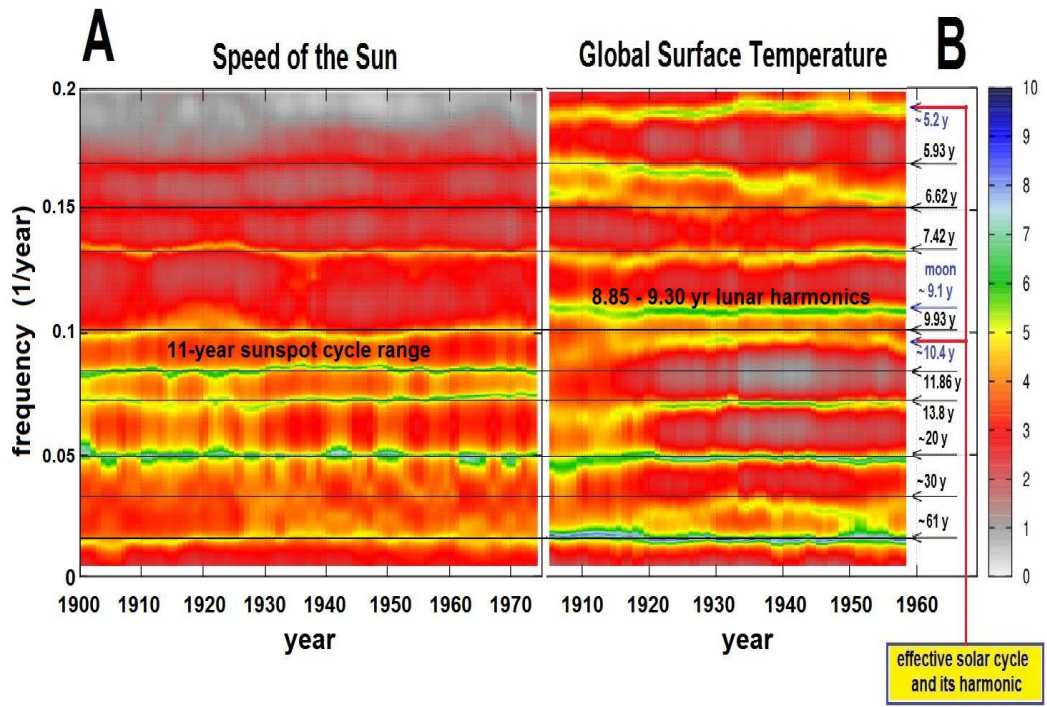}\protect\caption{{[}A{]} Time-frequency analysis ($L=110$ years) of the speed of the
Sun relative to the center of mass of the solar system. {[}B{]} Time-frequency
analysis ($L=110$ years) of the HadCRUT3 temperature record after
a quadratic fit was removed to eliminate the non stationary upward
bias \citep{Scafetta2010}. MEM-based spectrogram M = 660 (50\% of
the number of available monthly samples). The common astronomical
time-invariant spectral lines are highlighted with black straight
lines crossing the two panels. The blue arrow indicates the temperature
signature of the $\sim$9.1-year lunar tidal cycle, and the two red
arrow the effective temperature signatures of the solar cycle at $\sim$10.4
years and of its likely harmonic at $\sim$5.2 years. The panel colors
are in generic nonlinear units (to stress small peaks) from a minimum
power indicated as ``0'' to a maximum indicated as ``10''. The
spectral lines have an uncertainty of $\pm\text{\textonehalf}\nabla f=\pm\text{\textonehalf}/110=\pm0.0045$
$yr^{-1}$. Note the good spectral coherence at 20 and 60-year periods
that \citet{Holm} could not find (cf. Figure 3).}
\end{figure}

The $\sim$10.4-year temperature spectral line is slightly varying
between the 9.93-year and 11.86-year astronomical periods, which bound
the solar cycle dynamics as seen in Figure 2A \citep{Scafetta2012c}.
The period referring to the variable line at $\sim$10.4 years is
a little bit longer at the beginning of the record (up to about 11.86
years) and becomes shorter later (up to about 9.93 years). The evolution
of the $\sim$10.4-year oscillation fits the interpretation that this
oscillation is the temperature signature of the variable 11-year solar
oscillation. Indeed, from 1843 to 1913 the solar cycle was on average
11-12 years long while from 1913 to 2000 it was on average 10-11 years
long \citep[e.g.:][table 1]{Loehle,Thejll,Scafetta2012c}. See also
\citet{Scafetta2012c,Scafetta2012d} for more details about the relation
between the solar cycle and the Jupiter-Saturn cycles at 9.93 and
11.86 years, where it is also shown that the probability distribution
of the sunspot cycle length since 1750 presents two peaks with a major
peak around $10.3$ years and a secondary peak around $11.7$ years. 

There exists also a Venus-Earth-Jupiter model for the quasi 11-year
solar oscillation \citep{Hung,Scafetta2012c}. In this case the 11-year
solar cycle should be bounded between the 10.38-year and 11.68-year
average periods, which correspond to 16 and 19 times Venus-Jupiter
synodic cycle ($P_{VJ}=0.6486$ year), respectively. The varying temperature
line at 5.2 years could be a harmonic of the 10.4-year line and would
correspond to a recurrence made of 8 Venus-Jupiter synodic cycles,
which is not well observed in the solar speed record.

\subsection{The solar-lunar-astronomical origin of the 8\textendash 13 year temperature
oscillation band}

Let us explicitly clarify what happens in the 8-13 year period band
that according to Holm's time frequency analysis should be characterized
by a single variable line that breaks down in the interval 1920-1940
(see Figure 3B). This broken and variable pattern gave Holm the impression
that \citet{Scafetta2010}'s hypothesis is wrong. 

\citet{Scafetta2010,Scafetta2012b,Scafetta2013b} claimed that the
8-13 year temperature range is mostly regulated by two main astronomical
oscillations that, indeed, are quite visible in Figures 7B and 8B:
(1) a major solar-lunar tidal oscillation varying between 8.85 years
(the period of the recession of the line of the lunar apsides) and
9.3 years (the second harmonics of the lunar nodal cycle), the mean
is about 9.1 years \citep[ cf.][]{Wang2012}; (2) a variable 10-12
year Schwabe solar cycle. The temperature spectral peak at $\sim$10.4
years is approximately the average between the 9.93-year Jupiter-Saturn
spring tidal cycle and the 10.87 central sunspot frequency peak and
approximately corresponds to the larger peak of the bimodal solar
cycle length distribution \citet[figure 2]{Scafetta2012c}. 

Spectral analysis should advantage the solar cycles in the 10-11 year
period range over those in the 11-12 year period range because during
the 20th century most solar cycles were short and the Sun is likely
more active and the solar cycle amplitude larger during periods characterized
by cycles with a short length \citep[cf.:][]{Loehle,Thejll}. 

\begin{figure}[!t]
\centering{}\includegraphics[angle=-90,width=0.8\columnwidth]{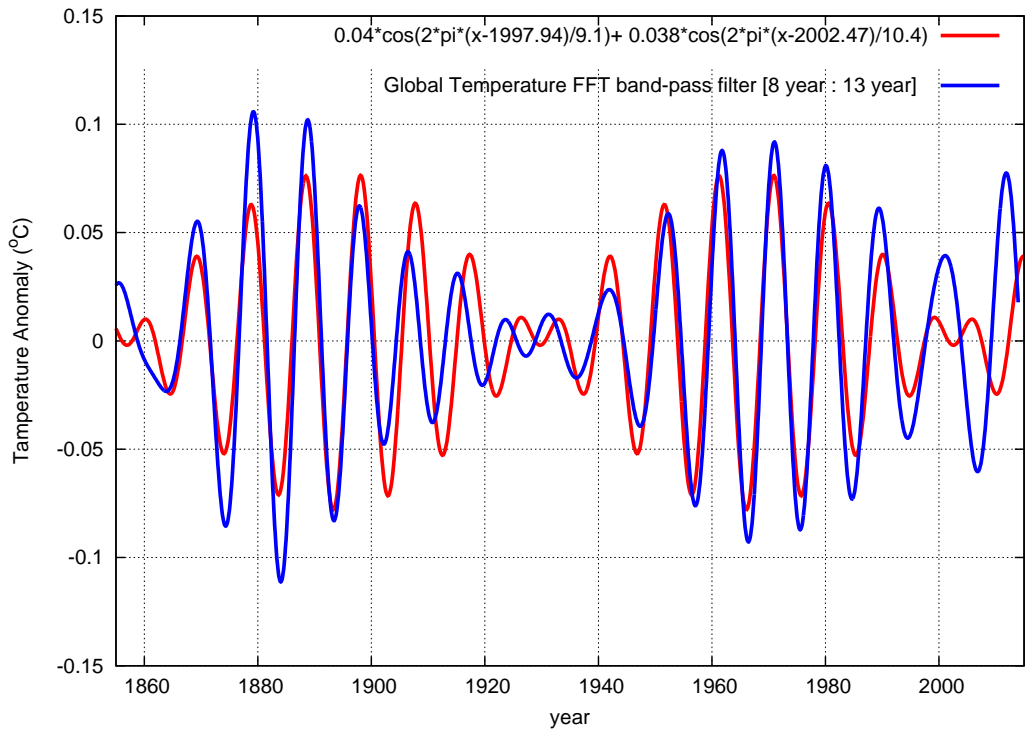}\protect\caption{8\textendash 13 year FFT band pass filter 8\textendash 13 of the temperature
record (blue) and Eq. \ref{eq:2} (red). The filtering highlights
that this frequency range is characterized by a strong beat with a
period of about 80 years.}
\end{figure}

Figure 9 shows a FFT band pass filter of the HadCRUT3 record (blue
curve) that isolates the global surface temperature variability between
8 and 13 year periods. This record clearly demonstrates the presence
of a beat oscillation with a period of about 70-90 years. Thus, at
least two major interfering harmonics determine the temperature variability
within this range and, according Eq. \ref{eq:beat}, spectral analysis
needs more than 80 years of data to separate them. This evidence again
indicates that Holm's chosen window lengths ($L=20$ to 60 years)
were too short and, therefore, inadequate to properly detect the constituent
harmonics of the temperature signal. 

In particular, Figure 9 clearly suggests that the two harmonics interfere
destructively from 1920 to 1940, when the time-frequency analysis
implemented by \citet[figure 2]{Holm}, and reported here in Figure
3B, shows a vanishing pattern. 

Because spectral analysis suggests that the two constituent harmonics
have average periods of 9.1 and 10.4 years, in first approximation
the modulation shown in Figure 9 can be modeled using two harmonics
as:

\begin{equation}
f(t)=0.040\cos\left(2\pi\frac{t-1997.94}{9.1}\right)+0.038\cos\left(2\pi\frac{t-2002.47}{10.4}\right).\label{eq:2}
\end{equation}
Figure 9 shows that the regression model, $f(t)$, (red curve) well
reproduces the filtered temperature curve (blue curve). 

The claim that these two global surface temperature oscillations likely
have an astronomical origin is based on two independent observations:
(1) the two frequencies match two major astronomical frequency ranges,
as explained above; (2) the phases of these two oscillations approximately
match the astronomical expectations, as explained below.

In fact, according Eq. \ref{eq:2} the 9.1-year oscillation peaks
in 1997.97. During this period the moon was crossing a nodal point
at the equinoxes \citep[supplement, page  35-36]{Scafetta2012b}.
Moreover, in 1997 the Moon was in its minor declination standstill:
a lunar declination standstill occurs every 9.3 years (\href{http://en.wikipedia.org/wiki/Lunar_standstill}{http://en.wikipedia.org/wiki/Lunar\_{}standstill}).
Solar and lunar eclipses were occurring in March and September: solar
eclipses occurred on 9/Mar/1997, 2/Sep/1997 and 26/Feb/1998; lunar
eclipses occurred on 24/Mar/1997, 16/Sep/1997 and 13/Mar/1998 (\href{http://www.timeanddate.com/eclipse/list.html}{http://www.timeanddate.com/eclipse/list.html}).
Therefore, the soli-lunar tidal torque was at its maximum because
mostly acting on the equator. Probably tidal forces could drive more
hot water from the equator toward the poles inducing a maximum in
the $\sim9.1$-year global surface temperature oscillation. A climatic
effect of the tidal forces moving the ocean water at different latitudes
may be relatively quick.

According Eq. \ref{eq:2} the 10.4-year temperature oscillation peaks
in 2002.47 \citep{Scafetta2012c}. Several solar indexes, such as
total solar irradiance, peaked around 2002 \citep[e.g.: ][]{ScafettaW2013,ScafettaW2014}
and the Jupiter-Saturn conjunction occurred in 2000.475 \citep{Scafetta2012c}.
Therefore, the maximum of the combination of the two astronomical
oscillations can be theoretically expected to have occurred between
2001 and 2002. A $\sim$1-year time-lag between the decadal solar/astronomical
oscillation and the equivalent temperature oscillation that peaked
around 2002.5 is physically plausible because of the thermal heat
capacity of the climate system \citep{Wigley}. A more detailed analysis
is left to another dedicated study.

\subsection{MVDR magnitude squared coherence (MSC) analysis}

As explained in Section 3.6, because of the shortness of the temperature
data the best coherence test can be made by directly comparing the
frequencies of the temperature spectrum with the theoretical astronomical
frequencies, as reported in Table 1. However, magnitude squared coherence
(MSC) analysis with $L=20$ and 30 year window was used by \citet{Holm}
to claim that the expected coherence could not be found: see Figure
3C. Here I discuss the problems of Holm's analysis and show how MSC
should be used for getting some reasonable results. 

Here I calculate the magnitude squared coherence (MSC) analysis using
the Capon\textquoteright s approach known as the \textit{minimum variance
distortion-less response} (MVDR) method \citep{Benesty}. The MVDR-MSC
method is chosen because it is a high-resolution method of coherence
analysis. It provides much sharper and reliable results than the one
based on the popular Welch\textquoteright s method implemented in
the MATLAB function \texttt{mscohere.m} used in \citet{Holm} and
reported here in Figure 3C. The MVDR-MSC method produces sharp peaks
approximately like MEM and it is more appropriate to separate close
frequencies using short windows. MVDR-MSC is based on the evaluation
of the following MSC equation:

\begin{equation}
\gamma_{xy}^{2}(\omega)=\frac{\left|S_{xy}(\omega)\right|^{2}}{S_{xx}(\omega)S_{yy}(\omega)}=\frac{\left|\mathsf{\mathbf{f^{\mathrm{\mathit{H}}}R_{\mathrm{\mathit{xx}}}^{\mathrm{\mathit{-1}}}R_{\mathrm{\mathit{xy}}}R_{\mathrm{\mathit{yy}}}^{\mathrm{\mathit{-1}}}f}}\right|^{2}}{\left[\mathbf{f^{\mathrm{\mathit{H}}}R_{\mathit{\mathit{xx}}}^{\mathit{-1}}f}\right]\left[\mathbf{f^{\mathit{H}}R_{\mathit{yy}}^{\mathit{-1}}f}\right]}\label{eq:4}
\end{equation}
where $S(\omega)$ is the cross-spectrum and $\mathrm{\mathbf{R}}$
is the cross-correlation ($L\times L$) matrix between the input time
series $x(t)$ and $y(t)$, and $\mathbf{f}$ is a vector made of
the harmonics of $\omega$, $f_{j}(\omega)=e^{i\omega j}/\sqrt{L}$,
with $j=0,1,\ldots L-1$, where $L$ is the window length. By mathematical
construction $0\leq\gamma_{xy}^{2}(\omega)\leq1,$ and $\gamma_{xy}^{2}(\omega)$
theoretically approaches 1 if the two original sequences present a
common major harmonic at the frequency $\omega$. See \citet{Benesty}
for details and computer simulations that demonstrate the superiority
of this methodology over the \texttt{mscohere.m} algorithm.

\begin{figure}[!t]
\centering{}\includegraphics[width=0.8\textwidth]{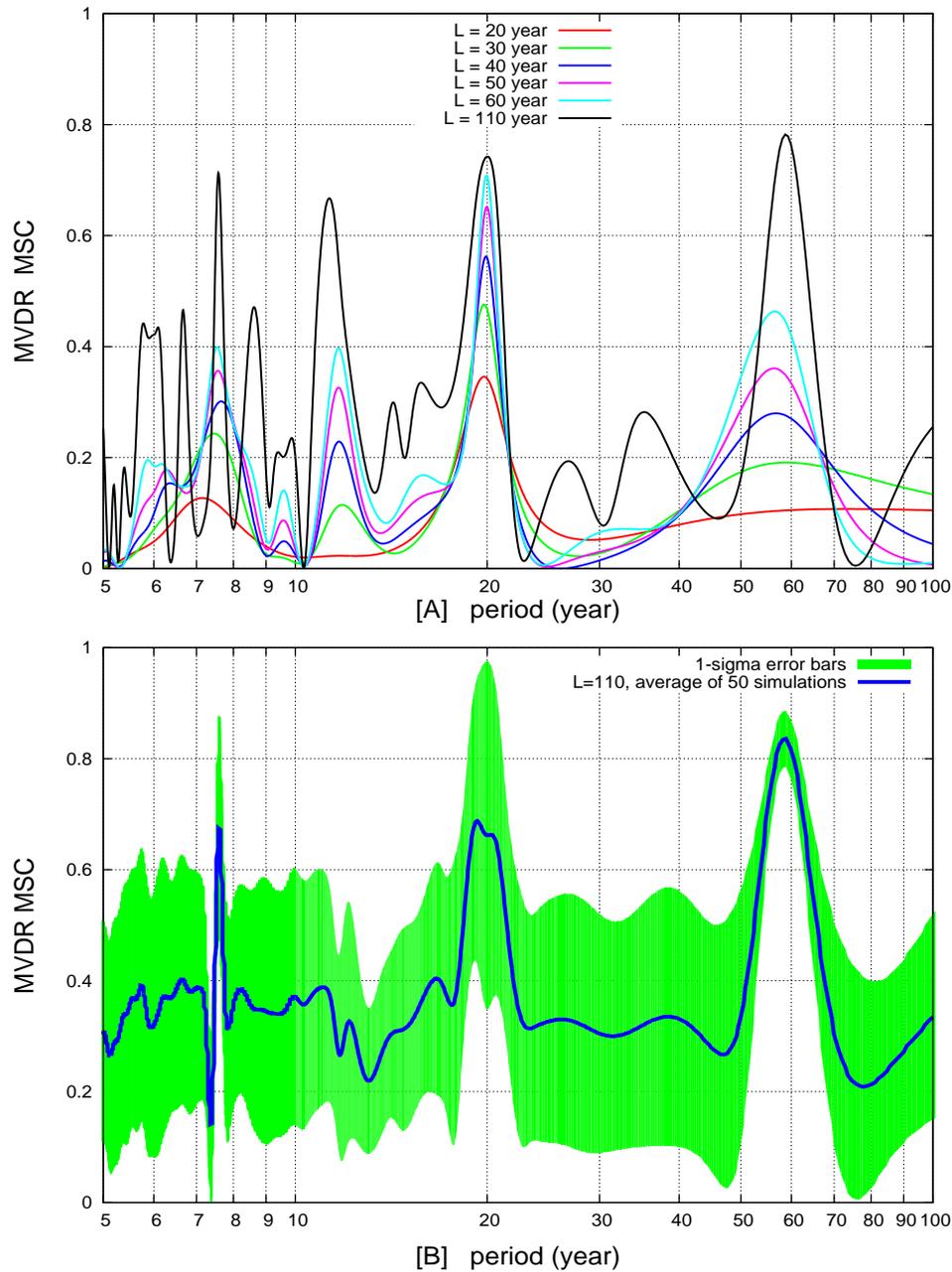}\protect\caption{{[}A{]} MVDR magnitude squared coherence (MSC) between the global
surface temperature record and the Sun's speed relative to the barycenter
of the solar system using Eq. \ref{eq:4}. Different window lengths
$L$ are used. The global surface temperature record is detrended
of its quadratic fit to eliminate the non stationary upward bias,
while the Sun's speed record is detrended only of its average value
before applying the MVDR MSC algorithm \citep{Benesty}. {[}B{]} Montecarlo
confidence test. It shows the average of 50 MVDR-MSC simulations between
the astronomical record and the synthetic records produced with Eq.
\ref{eq: 8-1}. MVDR-MSC peaks larger than 0.6-0.7 are statistically
significant.}
\end{figure}

Figure 10A shows the MVDR-MSC using $L=20$ and 30 year (I added the
curves for 40, 50 and 60 years because also tested but not shown by
Holm), and $L=110$ years that should be chosen to better solve the
close major harmonics that can be detected in the 164-year temperature
record. Figure 10A clearly shows that by increasing the window length
to $L=110$ years a larger number of expected coherence peaks become
manifest. Four major coherence peaks ($\gamma_{xy}^{2}(\omega)\gtrapprox0.7$)
appear at about 7.5, 11, 20 and 60 year periods. Minor coherence peaks
at ($0.4<\gamma_{xy}^{2}(\omega)<0.5$) may exist also at about 6,
6.7 and 8.7 periods, but this result is less certain. 

Note that a sharp coherence peak within the range 8.85-9.30 year is
not supposed to be seen in Figure 10A because this period refers to
lunar harmonics not present in the Sun's speed. If the temperature
oscillations present a small degree of non-linearity and chaotic variation,
as reasonable, its MSC values would be smaller than the case of perfectly
harmonic oscillations.

The result depicted in Figure 10A confirms what could be deduced visually
from Figures 5 and 7 (see Table 1) and Figure 8, and contradicts the
results of \citet{Holm} (depicted here in Figure 3C) who concluded
that only a weak ``coherence at a period of 15-17 years can be found
in the data.'' Indeed, Holm used a MSC methodology with windows of
length $L=20$ years and $L=30$ years, which are even shorter than
the already inadequate $L=60$ year window length used for his time-frequency
analysis: their spectral resolution is $\nabla f_{20}=1/20=0.05$
$yr^{-1}$ and $\nabla f_{30}=1/30=0.033$ $yr^{-1}$. Moreover, the
function \texttt{mscohere.m} used in \citeauthor{Holm} produces large
lobes and does not separate well close frequencies \citep{Benesty}. 

\citet{Benesty} gives no statistical properties for the MVDR-MSC
estimate. To test the significance of the coherence spectral peaks
shown in Figure 10A, a Montecarlo test is performed. The astronomical
record, whose harmonic dynamics is well defined and the record is
highly accurate, is processed against an ensemble of synthetic temperature
signals made of the three proposed coherence harmonics of the type:

\begin{equation}
g(t)=\cos\left(\frac{2\pi t}{7.5+0.17\vartheta}+\phi\right)+1.7\cos\left(\frac{2\pi t}{20+1.2\vartheta}+\phi\right)+4.3\cos\left(\frac{2\pi t}{60+5\vartheta}+\phi\right)+\xi_{t}\label{eq: 8-1}
\end{equation}
where $\phi$, $\vartheta$ and $\xi_{t}$ are Gaussian normal random
noise. The relative amplitudes of Eq. \ref{eq: 8-1} were deduced
from Figure 7B in units of the smallest one among the three. The periods
were made to vary within a reasonable spectral error deduced from
Table 1 and Figure 7B to simulate possible nonlinear and chaotic climatic
responses to the hypothesized external harmonic forcing. The purpose
of the test was to study the statistics of the three expected coherent
MVDR-MSC peaks. 

Fifty simulations were run. Their average and standard deviation band
are shown in Figure 10B. The simulation produces the expected coherence
peaks at 7.5, 20 and 60 year periods with average MVDR-MSC values
between 0.65 and 0.85, which is a range compatible with the peak values
observed in Figure 10A. Thus, limited to the three chosen harmonics
and relative to the MVDR-MSC measure, the temperature signal is statistically
equivalent to the harmonic model of Eq. \ref{eq: 8-1}. Figure 10B
also shows that using a 164-year monthly record MVDR-MSC values smaller
than 0.6 would not be distinguishable from noise.

Finally, the inadequacy of Holm's analysis can be easily demonstrated
using computer generated sequences. I generated two artificial sequences
of 164 data points, like the 164-year temperature signal, of the type 

\begin{equation}
f(t)=\sum_{i=1}^{6}\cos(2\pi\Omega_{i}t+\phi_{i})+\xi_{t}\label{eq: 8}
\end{equation}
where $\phi_{i}$ are random phases, $\xi_{t}$ is Gaussian normal
random noise, and the frequencies $\Omega_{i}$ are similar to those
found in the temperature record: $\Omega_{1}=1/60$ $yr^{-1}$; $\Omega_{2}=1/20$
$yr^{-1}$; $\Omega_{3}=1/10.4$ $yr^{-1}$; $\Omega_{4}=1/9.1$ $yr^{-1}$;
$\Omega_{5}=1/7.5$ $yr^{-1}$; $\Omega_{6}=1/6.5$ $yr^{-1}$. Then
I calculated the MSC between the two generated signals using the MATLAB
function \texttt{mscohere.m} as used in \citet{Holm} with windows
of $L=20$ and 30 years, and I repeat the analysis of the same computer
generated records using the MVDR MSC method with $L=110$ years. The
results are depicted in Figure 11. Only in the low panel case (blue
curve), which uses the MVDR MSC method, the 6 common frequencies are
clearly detected. On the contrary, the two upper panels show vague
\texttt{mscohere} curves made of two wide lobes within the 5-10 and
10-20 year period bands: this is the same pattern shown in Holm's
figures 4 and 5 (his figure 5 is shown here in Figure 3C). Thus, the
evidence is that Holm's MSC methodology fails  to identify close harmonics
similar to those that need to be searched and, therefore, it cannot
be used to test whether or not the temperature signal contains astronomical
harmonics. 

\begin{figure}[!t]

\begin{centering}
\includegraphics[width=0.8\textwidth]{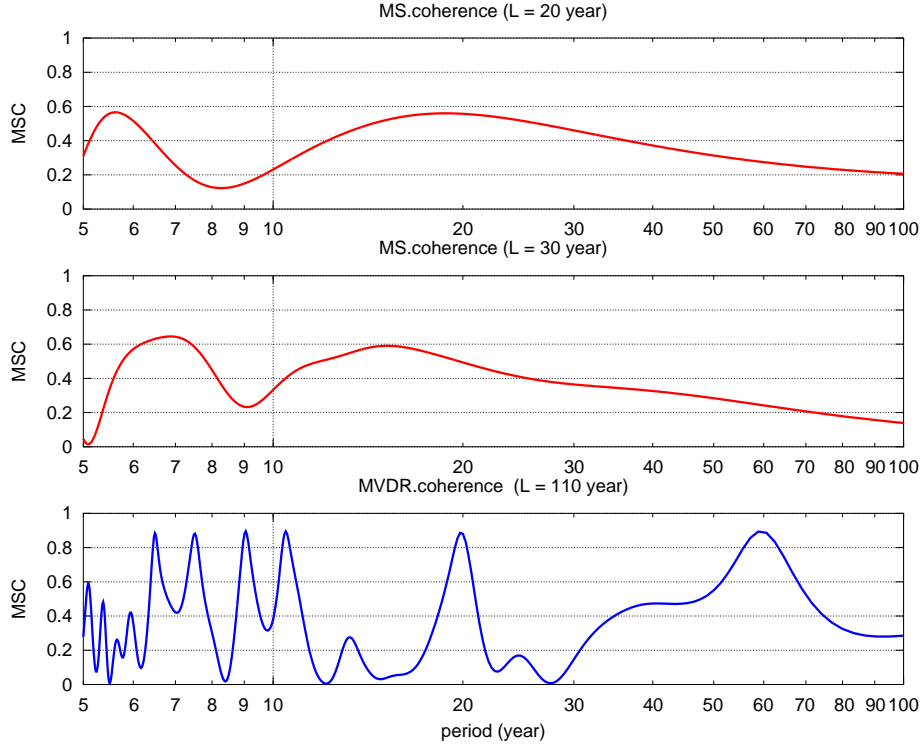}\protect\caption{Typical magnitude squared coherence (MSC) between two computer generated
signals made of 6 harmonics plus random noise (Eq. \ref{eq: 8}).
The upper two panels use the MATLAB function \texttt{mscohere.m} adopted
in \citet{Holm} with windows of $L=20$ and 30 years (red curves).
The bottom panel use the MVDR coherence method with $L=110$ years
(blue curves). Only in the latter case the 6 theoretical frequencies
($\Omega_{1}=1/60$ $yr^{-1}$; $\Omega_{2}=1/20$ $yr^{-1}$; $\Omega_{3}=1/10.4$
$yr^{-1}$; $\Omega_{4}=1/9.1$ $yr^{-1}$; $\Omega_{5}=1/7.5$ $yr^{-1}$;
$\Omega_{6}=1/6.5$ $yr^{-1}$) are well detected. }

\par\end{centering}

\end{figure}

In conclusion, Holm's analysis, by using too short spectral windows
and a poor MSC algorithm, is corrupted by too large frequency error
bars that within the analyzed scales cover contiguous expected astronomical
peaks. This is suggested by the two wide spectral lobes shown in Figure
3C, which are reproduced in the computer experiment depicted in Figure
11. With short $L=20$ and 30 year windows the \texttt{mscohere.m}
algorithm simply clusters various contiguous harmonics into two average
groups, one between 5 and 10 years and the other between 10 and 20
years; the 60-year coherence cannot be detected at all because such
a period simply does not enter within a 20 or 30 year window. Nevertheless,
Holm plotted his coherence graphs (e.g. Figure 3C) up to periods of
100 years giving the misleading impression that there is no spectral
coherence at the 60-year period.

\section{Conclusion}

Since 1859, when \citeauthor{Wolf} proposed that solar variability
could be modulated by the combined effect of the planets, solar scientists
have been wondering about the possibility of a planetary theory of
solar variation, which would also imply a planetary theory of climate
change. Numerous papers have recently appeared in the scientific literature
supporting such a theory using empirical analysis of data and proposing
models that attempt to interpret the data patterns with some combinations
of planetary harmonics \citep[e.g.:][and many others]{Abreu,Jakubcova,McCracken2014,McCracken2013,Morner,Puetz,Salvador,Scafetta2010,Scafetta2012d,Scafetta2013a,Scafetta2013b,Scafetta2014,ScafettaW2013,Scafettaw2013a,ScafettaW2014,Tan,Wilson}.
However, some critiques have also appeared questioning the empirical
findings.

I have shown that \citet{Cauquoin} used interesting data and analysis
but they misinterpreted the complexity of the planetary theory of
solar variation because they focused only on the planetary model proposed
by \citet{Abreu} and did not highlight some important details that
their analysis was revealing. I have shown that \citet{Cauquoin}'s
own analysis of high-resolution cosmogenic solar activity proxy records
both during the Holocene and during the Marine Interglacial Stage
9.3 (MIS 9.3), 325\textendash 336 kyear ago reveals that these two
independent solar proxy records show four common spectral peaks at
about 103, 115, 130 and 150 years at about the 95\% or above confidence
level. These are the typical 100-150 year oscillations that regulate
the alternating of grand solar minima and maxima similar to the Wolf,
Sp\"orer, Maunder and Dalton grand solar minima observed during the
last centuries. I have shown that the planetary model proposed by
\citet{Scafetta2012c} predicts these four harmonics well. The peculiarity
of my model over that proposed by \citet{Abreu} is that in my model
solar variation is due to a modulation of the Schwabe 11-year solar
cycle by planetary harmonic tidal forcings plus a reasonable non-linear
response of the internal solar mechanisms to the external planetary
harmonic forcings. Therefore, solar variation depends on a coupling
between external and internal solar mechanisms. This model may be
physically more plausible to explain some of the typical observed
harmonics found in the solar proxy models \citep[cf. ][]{Ogurtsov}.

On the decadal and multidecadal scale \citet{Holm} claimed that time-frequency
analysis based on $L=60$ year window and magnitude squared coherence
(MSC) analysis based on $L=20$ and 30 year window would demonstrate
that global surface temperature data do not contain time-invariant
spectral lines nor frequencies coherent to astronomical oscillations.
He questioned \citet{Scafetta2010} where it was first proposed that
the climate system contains a signature of multiple astronomical harmonics
of solar, lunar and planetary origin. I have demonstrated that \citet{Holm}
conclusion is flawed because based on a misapplication of the time-frequency
analysis methodology that also yield astronomical paradoxes. Essentially,
the windows lengths used in \citet{Holm} are too short to separate
the frequencies of interest. Using window lengths of $L=60$ years
for the time frequency analysis, the spectral resolution of his analysis
is $\nabla f_{60}=1/60=0.0167$ $yr^{-1}$; using window lengths of
$L=20$ and 30 year for the magnitude squared coherence analysis,
the spectral resolution of his analysis is $\nabla f_{20}=1/20=0.05$
$yr^{-1}$ and $\nabla f_{30}=1/30=0.033$ $yr^{-1}$, respectively.
However, the theoretical spectral resolution required for separating
the astronomical expected harmonics is $\nabla f\leq1/178.34=0.0056$
$yr^{-1}$, as seen in Figure 5B. Therefore, Holm's methodologies
did not have the required spectral resolution to do the job and could
not test whether or not the global surface temperature contains a
signature of astronomical frequencies.

I have demonstrated that only a minimal windows of $L\gtrapprox178.4$
years can theoretically well solve the beats among the expected major
astronomical frequencies, but the temperature record is only 164 years
long. Therefore, the best analysis that can be done right now is to
use only one 164-year window covering the entire available record
as done in \citet{Scafetta2010} and repeated here in Table 1 by directly
comparing the spectra of Figures 5 and 7B with their confidence error.
Table 1 shows that a spectral coherence among the global surface temperature
and the main astronomical expected oscillations is clearly detected. 

However, even if partially, a minimal window length of $L=110$ years
can solve at least the major expected harmonics. Using $L=110$ year
both the time-frequency analysis and the magnitude squared coherence
analysis clearly reveal the spectral coherence among the main global
surface temperature and the astronomical oscillations that can be
detected using the available 164-year long global surface temperature
sequences, as shown in Figures 8 and 10. Finally, Figures 6 and 11
demonstrate the inadequacy of Holm's analysis with simple computer
experiments. 

Indeed, several climatic oscillations have been observed throughout
the Holocene and can be easily associated with solar, planetary and
lunar harmonics as discussed in the Introduction.

In conclusion, \citet{Cauquoin} is quite interesting but these authors
misunderstood the complexity of the planetary theory of solar variation.
On the contrary, \citet{Holm}'s results are definitely artifacts
of his inadequate window size used in both his time frequency analysis
and magnitude squared coherence analysis, which was even worsen by
the low quality \texttt{mscohere.m} algorithm that he used: essentially,
his results are misleading.

\section*{Appendix}

\subsection*{A) The three frequency solar model}

Here I summarize the functions used for constructing the planetary/solar
three frequency solar model discussed in Section 2 and shown in Figure
2. This appendix reproduces the Appendix in \citet{Scafetta2012c}
for the reader convenience.

The three basic proposed harmonics are:

\begin{equation}\label{}     
h_1(t)=0.83~\cos\left(2\pi ~\frac{t-2000.475}{9.929656}\right) 
\end{equation}

\begin{equation}\label{}     
h_2(t)=1.0~\cos\left(2\pi ~\frac{t-2002.364}{10.87}\right) 
\end{equation}

\begin{equation}\label{}     
h_3(t)=0.55~\cos\left(2\pi~\frac{t-1999.381}{11.862242}\right)~, \end{equation} 
where the relative amplitudes are weighted on the sunspot number record since 1749. Three frequencies derive from the spectrum of the sunspot record (see Figure 2A) where the two side harmonics at 9.93 and 11.86 year period are theoretically deduced from the tidal oscillations generated by Jupiter and Saturn. The three phases are deduced: from the conjunction date of Jupiter and Saturn, $t=2000.475$; the perihelion date of Jupiter, $t=1999.381$; and by regression on the sunspot cycle, $t=2002.634$.

The basic harmonic model is
\begin{equation}\label{Eq13}     
h_{123}(t)=h_1(t)+h_2(t)+h_3(t) 
\end{equation}

\begin{eqnarray}\label{Eq14}  
 f_{123}(t)= h_{123}(t) & \qquad if & \qquad h_{123}(t)\geq 0 \\  
f_{123}(t)= 0&  \qquad if & \qquad h_{123}(t)<0
\end{eqnarray} 

which is depicted in Figure 2B. 

The chosen beat function modulations in generic relative units and their sum are:

\begin{equation}\label{Eq15}     
b_{12}(t)=0.60~\cos\left(2\pi~\frac{t-1980.528}{114.783}\right) \end{equation}

\begin{equation}\label{}     
b_{13}(t)=0.40~\cos\left(2\pi~\frac{t-2067.044}{60.9484}\right) \end{equation}

\begin{equation}\label{}     
b_{23}(t)=0.45~\cos\left(2\pi~\frac{t-2035.043}{129.951}\right) \end{equation}

\begin{equation}\label{Eq18}     b_{123}(t)=b_{12}(t)+b_{13}(t)+b_{23}(t)+1 ~. 
\end{equation} 
The three relative amplitudes are roughly estimated against Eq. \ref{Eq14}. The millennial modulating function is

\begin{equation}\label{eq19}     
g_m(t)=A~\cos\left(2\pi~\frac{t-2059.686}{983.401}\right)+B 
\end{equation}
The parameters $A$ and $B$ may be changed according to the application. The two proposed modulated solar/planetary functions are

\begin{equation}\label{eq20}     F_{123}(t)=g_m(t)~f_{123}(t)~~~\texttt{with}~~A=0.2, B=0.8 \end{equation}
\begin{equation}\label{Eq21}     B_{123}(t)=g_m(t)~b_{123}(t)~~~\texttt{with}~~A=0.3,B=0.7~. \end{equation} 

See \citet{Scafetta2012c} for more details and for a supplement file
with the actual data.

\subsection*{B) \citet{Scafetta2010} astronomical - temperature spectral coherence}

For convenience of the reader Figure 12 reproduces figure 6B and 9A
of \citet{Scafetta2010}. Figure 12B shows MEM evaluations of numerous
climatic records such as the global surface temperature (G), the northern
and southern global surface temperatures (GN and GS), the global,
northern and southern land surface temperatures (L, LN, LS) and the
global, northern and southern ocean surface temperatures (O, ON, OS).
The green bars are the main solar, astronomical and lunar expected
harmonics (cf. Figure 5). It is easy to notice a coherence between
the astronomical harmonics and the MEM spectral peaks at multiple
frequencies. Figure 12A directly compares the temperature average
periods (red) against the astronomical average periods (blue). A $\chi^{2}$
test output among the various frequencies is shown suggesting that
the coherence confidence is at the 96\%. Additional calculations and
evidences are provided in \citet{Scafetta2010}.

\begin{figure}[!t]

\centering{}\includegraphics[width=0.8\textwidth]{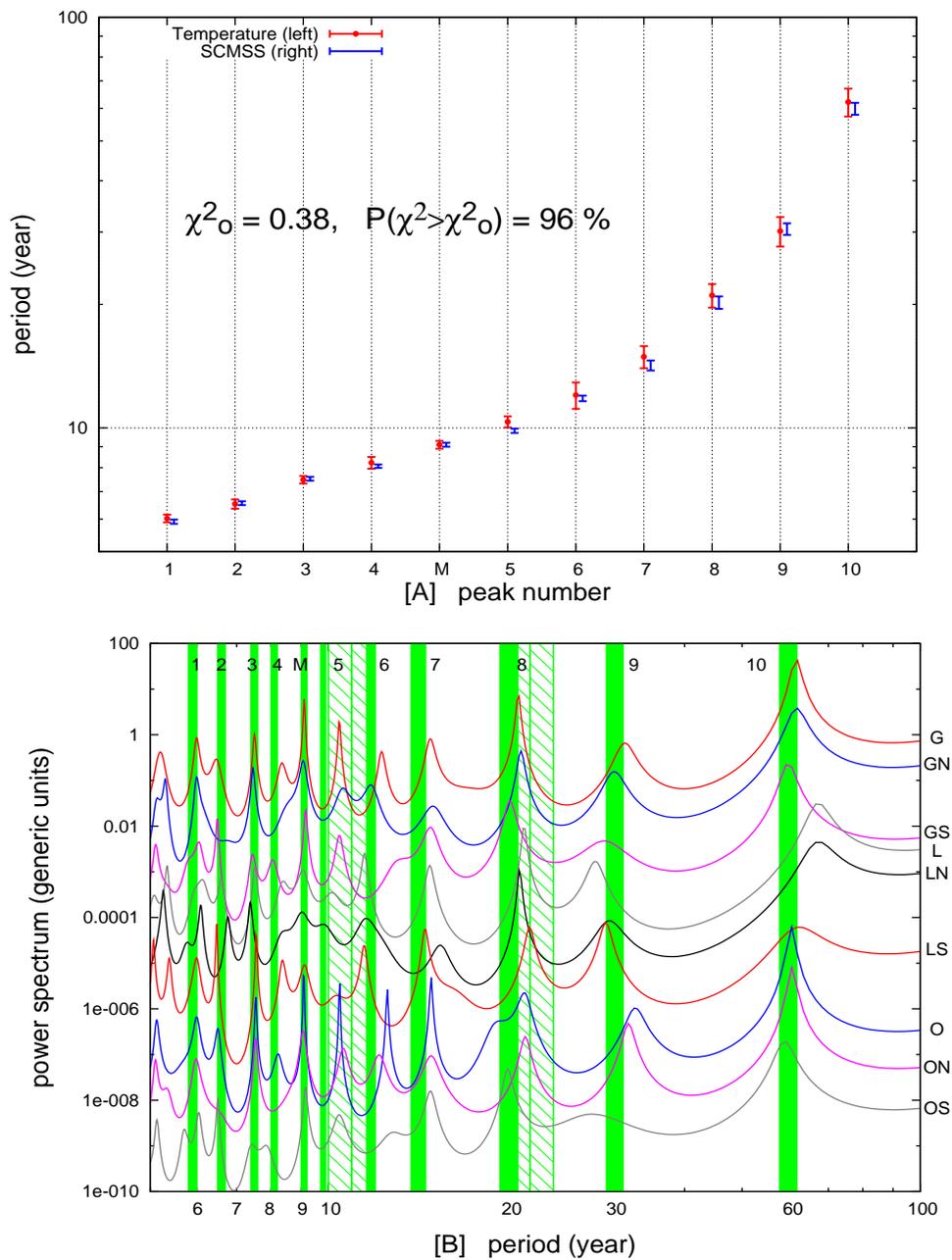}\protect\caption{Reproduction of figure 6B and 9A of \citet{Scafetta2010} showing
{[}A{]} the $\chi^{2}$ spectral coherence test and {[}B{]} the direct
comparison between the MEM curve of several climatic records and the
astronomical, solar and lunar harmonics (green bars). See the original
paper for details and addition.}
\end{figure}

\subsection*{C) Kepler's diagram of Jupiter-Saturn conjunctions}

Figure 13 shows the original diagram of Jupiter-Saturn conjunctions
prepared by \citet{Kepler}. It highlights the date and the constellation
position of the great conjunctions, that occur every 20 years, from
1583 to 1763. The 60-year trigon pattern, that involves three consecutive
conjunctions, is clearly visible together with its slow millennial
rotation. The 20, 60 and 800-1000 year oscillations associated to
the movement of Jupiter and Saturn were well known since antiquity
and used to construct some kind of astrological-based climate models
\citep{Kepler,Iyengar,Masar,Temple}. See \citet{Scafetta2012a} for
additional details.

\begin{figure}[!t]
\begin{centering}
\includegraphics[width=0.8\textwidth]{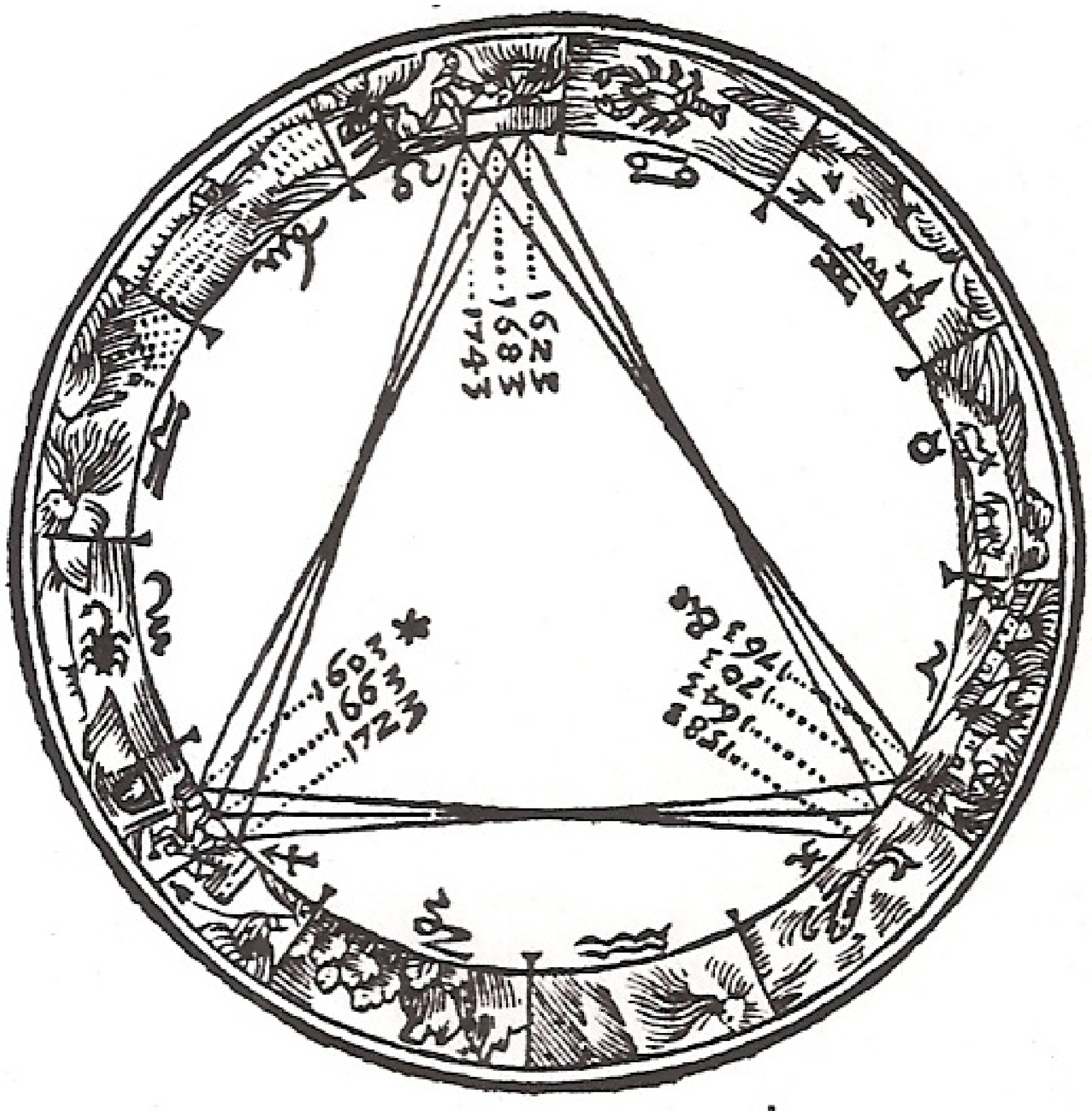}\protect\caption{The original diagram of Jupiter-Saturn conjunctions prepared by \citet{Kepler}}

\par\end{centering}

\end{figure}

\newpage{}

\end{document}